\documentclass[a4paper,12pt]{article}

\usepackage{url}
\usepackage{epsfig}
\usepackage{amsmath}
\usepackage{amssymb}
\usepackage{amsfonts}
\usepackage{bbm}
\usepackage[footnotesize]{caption}
\usepackage{graphicx}
\usepackage{mathrsfs}
\usepackage[font=scriptsize]{subcaption}
\usepackage{color}
\usepackage[dvipsnames]{xcolor}
\usepackage{comment}
\usepackage{mathtools}
\usepackage{braket}
\usepackage{cite}
\usepackage{hyperref}
\usepackage{multirow}
\usepackage{relsize}
\usepackage{fullpage}
\usepackage{makecell}
\usepackage{enumitem,xcolor}
\usepackage{doi}

\begin{document}

\begin{titlepage}
\begin{center}
{\bf\Large   Exceptional Unification of Families and Forces} \\[12mm]
Alfredo Aranda$^{\ddag}$%
\footnote{E-mail: \texttt{fefo@ucol.mx}},
Francisco~J.~de~Anda$^{\dagger}$%
\footnote{E-mail: \texttt{fran@tepaits.mx}},
Stephen~F.~King$^{\star}$%
\footnote{E-mail: {\tt king@soton.ac.uk}; ORCID: https://orcid.org/0000-0002-4351-7507},
\\[-2mm]

\end{center}
\vspace*{0.50cm}
\centerline{$^{\ddag}$ \it
Facultad de Ciencias-CUICBAS, Universidad de Colima, C.P.28045, Colima, M\'exico 01000, M\'exico}
\centerline{Dual CP Institute of High Energy Physics, C.P. 28045, Colima, M\'exico}
\vspace*{0.2cm}
\centerline{$^{\dagger}$ \it
Tepatitl{\'a}n's Institute for Theoretical Studies, C.P. 47600, Jalisco, M{\'e}xico}
\vspace*{.20cm}
\centerline{$^{\star}$ \it
School of Physics and Astronomy, University of Southampton,}
\centerline{\it
SO17 1BJ Southampton, United Kingdom }
\vspace*{1.20cm}

\begin{abstract}
{\noindent
This work considers the remarkable suggestion that the three families of quarks and leptons may be unified, together 
with the Higgs and gauge fields of the Standard Model (SM), into a single ``particle'', namely the 
$\textbf{248}$ vector superfield of 
a ten-dimensional $E_8$ super Yang Mills (SYM) theory.
Towards a realistic model along these lines,
a class of orbifoldings based on 
$T^6/(\mathbb{Z}_N\times \mathbb{Z}_M)$ are proposed and explored, that can in principle break $E_8$ SYM 
down to the minimal supersymmetric standard model (MSSM), embedded in a larger group such as $E_6$, $SO(10)$ or $SU(5)$, together with other gauge group factors which can be broken by Wilson lines.
A realistic model based on $T^6/(\mathbb{Z}_6\times \mathbb{Z}_2)$ is presented.  
The orbifold breaks $E_8$ SYM down to 
a Pati-Salam gauge group in 4d, together with other gauge groups, 
which are subsequently broken to 
the SM gauge group with proto-realistic fermion mass matrices,
and experimental signals associated with a low Pati-Salam scale.}
\end{abstract}
\end{titlepage}

\section{Introduction}
\label{sec:intro}
Grand Unified Theories (GUTs) unify the three independent gauge interactions of the Standard Model (SM) 
gauge group $G_{321}$ into a larger 
gauge symmetry~\cite{Pati:1973uk, Georgi:1974sy, Fritzsch:1974nn, Mohapatra:1974gc, Pati:1974yy, Georgi:1979dq}. 
GUTs also unify the representations of fermions (and scalars) into a smaller number of 
simpler ones. For example, the SM gauge group may be unified into a simple $SU(5)$ gauge group, with 
quarks and leptons in three copies of the $\overline{\textbf{{5}}}$ and $\textbf{10}$ representations. The gauge group may be further enlarged to $SO(10)$ which unifies the SM fermions into three copies of the irreducible $\textbf{16}$ representation (predicting right-handed neutrinos). The even larger group $E_6$ contains the previous features and extends them by including the Higgs in the same representation - provided one has $\mathcal{N}=1$ supersymmetry (SUSY). 

The sequence of unified groups may be further extended to include the exceptional groups $E_{6,7,8}$, familiar from the Dynkin diagram analysis of Lie groups~\cite{Buchmuller:1985rc,Koca:1982zi},
\begin{equation}
G_{321}\equiv SU(3)_C\times SU(2)_L\times U(1)_Y \subset SU(5) \subset SO(10) \subset E_6 \subset E_7 \subset E_8,
\end{equation}
where $E_8$ is the largest finite exceptional Lie group.
$E_{7}$ and $E_8$ are not normally used in GUTs. The main reason for this is that they do not have complex representations, which implies on the one hand, the presence of currently unobserved (and tightly constrained) mirror fermions, and on the other, the necessity of {\it separating} them to obtain the chiral fermion structure of the SM, where such separation is highly non-trivial. In spite of this general situation, interest in these groups as potential GUTs exists due in no small measure by the fact that they contain the following subgroup structure:
\begin{equation}
SU(5)\times SU(3)_F \subset E_7, \ \ {\rm and} \ \ SO(10)\times SU(3)_F \subset E_8,
\end{equation}
where $SU(3)_F$ may be identified as a ``family'' or ``flavour'' symmetry. This flavour symmetry in the context of GUTs has been widely explored in the literature, within different unification settings. Such scenarios are sometimes referred to as ``Flavoured GUTs'' (possibly including  SUSY~\cite{King:2001uz,King:2017guk,Hagedorn:2010th,Antusch:2014poa,Bjorkeroth:2015ora,Bjorkeroth:2015uou,Bjorkeroth:2017ybg,deAnda:2017yeb} and/or extra-dimensions (ED)~\cite{Altarelli:2008bg,Burrows:2009pi,Burrows:2010wz,deAnda:2018oik,Altarelli:2006kg,Adulpravitchai:2010na,Adulpravitchai:2009id,Asaka:2001eh,deAnda:2019jxw,deAnda:2018ecu,deAnda:2018yfp}). Flavoured GUTs have also been proposed 
based on other groups such as $SU(7)$ \cite{Kang:1981nr,Cox:1980er,Hwang:2002hg}, $SU(8)$ \cite{Buchmuller:1983iu,Baaklini:1980ju}, $SU(11)$ \cite{Georgi:1979md,Kim:1979uh} $SO(16)$ \cite{Arnold:1985mv,Kawamura:2009gr}, $SO(18)$ \cite{Reig:2017nrz,Bagger:1984qh,Bagger:1984gz,Wilczek:1981iz}.
While there have been several models proposed based on $E_8$~\cite{Adler:2002yg,Adler:2004uj,Garibaldi:2016zgm,Thomas:1985be,Konshtein:1980km,Baaklini:1980fv,Baaklini:1980uq,Barr:1987pu,Bars:1980mb,Koca:1981xd,Mahapatra:1988gc,Ong:1984ej,Camargo-Molina:2016yqm}, 
none of them includes a complete unification of families, Higgs and gauge bosons into a single multiplet. 

The gauge group $E_8$ is particularly attractive from the point of view of unification 
since its adjoint $\textbf{248}$ representation is also the fundamental representation, and so all three families of fermions, 
together with Higgs scalars and gauge vector bosons may all lie in the same representation. This would be the ultimate unification: all matter, Higgs and gauge forces arising from one 
$E_8$ ``particle'', namely the $\textbf{248}$ representation, together with one $E_8$ gauge force.
Such a supermultiplet suggests an $\mathcal{N}=4$ super Yang Mills (SYM) theory in four dimensions (4d), commonly regarded as the simplest quantum field theory, which has the property of being completely finite. Unfortunately, $\mathcal{N}=4$ in 4d is very far from the SM, since both the gauge forces and the extra supersymmetries need to be broken somehow, and it is not clear how to do this if one starts from 4d.

One interesting suggestion, is to 
start from $\mathcal{N}=1\ \ E_8$ SYM in 10d, compactified using a coset space reduction which breaks both the $E_8$ and the would-be extended $\mathcal{N}=4$ SUSY in 4d, to the SM group with $\mathcal{N}=1$ SUSY in 4d, also removing the mirror fermions, as desired~\cite{Olive:1982ai}. The coset space reduction is achievable by orbifold compactification, and the resulting 4d effective theory would be the minimal supersymmetric standard model (MSSM).
These ideas appeared just before the first superstring revolution~\cite{Polchinski1}, leading to the heterotic superstring based on $E_8\times E_8$, followed by the second superstring revolution~\cite{Polchinski2}
leading to $M$-theory and $F$-theory, all of which promised new insights into gravity. Although many of these approaches also involve $E_8$ in 10d, the fundamental starting point is very different, namely superstrings and branes as the basic objects \cite{Parr:2020oar}, and the goals and objectives of these theories are very different, namely to relate (quantum) gravity to gauge theories in a unified structure.
Consequently, the proposal of family unification based on $\mathcal{N}=1\ \ E_8$ SYM in 10d was partially eclipsed by the superstring revolutions, and the original idea~\cite{Olive:1982ai} has been largely neglected. 
In particular a realistic model in which the three families of quarks and leptons are unified, together 
with the Higgs and gauge fields of the Standard Model (SM) into a single 
$\textbf{248}$ vector superfield of an $\mathcal{N}=1\ \ E_8$ SYM theory in 10d, was never developed.
This extraordinarily elegant hypothesis is worthy of further exploration, and the development of a realistic model along these lines is long overdue.

Towards a realistic model, this paper proposes and explores a class of orbifoldings based on 
$T^6/(\mathbb{Z}_N\times \mathbb{Z}_M)$, which can in principle break $E_8$ SYM 
down to the minimal supersymmetric standard model (MSSM), possibly embedded in a 
larger group such as $E_6$, $SO(10)$ or $SU(5)$, together with other gauge group factors which can in principle be 
broken using Wilson lines.
A promising example with  $T^6/(\mathbb{Z}_6\times \mathbb{Z}_2)$ is found that breaks $E_8$ SYM down to 
a Pati-Salam gauge group in 4d~\cite{Pati:1974yy}, together with other gauge groups, which are further broken to the SM
by Wilson lines in the right-handed neutrino directions, allowing proto-realistic fermion mass matrices.
Thus the whole SM field content, together with the forces, are unified in a single $E_8$ SYM field. 
However, it proves necessary to add additional multiplets for anomaly cancellation.
Although such states are expected to
gain large masses, they could play a role in unification.
It is important to emphasise that this work deals with a 
field theory based on point particles, and unlike apparently related string theory~\cite{Ibanez:1987pj} or F-theory models~\cite{King:2010mq}, 
gravity is not included in the present framework. On the other hand, gravity might not be a fundamental force of nature
and could arise as an emergent phenomenon~\cite{deAnda:2019tri,Linnemann:2017hdo,Barcelo:2001ah}.

The layout of the remainder of the paper is as follows.
In Section~\ref{sec:susy} the $E_8$ SYM theory is introduced,
the $\mathcal{N}=4$ SUSY Lagrangian in $R^4\times T^6$ is presented,
and it is shown how the extended SUSY may be broken by orbifolding.
Section~\ref{sec:e8}  discusses $E_8$ gauge breaking, 
first by considering the orbifolding $T^6/(\mathbb{Z}_N)$ which breaks the gauge group 
$E_8$ into different subgroups for different choices of $N$, then by considering 
a general $T^6/(\mathbb{Z}_N\times \mathbb{Z}_M)$ orbifolding, which preserves 
$\mathcal{N}=1$ SUSY, and finally by adding Wilson lines to break the rank of the gauge group.
In Section~\ref{examples} some examples of $E_8$ breaking for various values of $N,M$ are discussed,
including: $E_6\times SU(3)_{F}$ from $T^6/\mathbb{Z}_3$; 
E$_6$ from $T^6/(\mathbb{Z}_3\times \mathbb{Z}'_3)$;
$SO(10)$ and $SU(5)$ cases from $T^6/\mathbb{Z}_6$; and finally a report on a general search for the 
MSSM from $T^6/(\mathbb{Z}_N\times \mathbb{Z}_M)$ is given.
Section~\ref{sec:psorb} considers a Pati-Salam model from $T^6/(\mathbb{Z}_6\times \mathbb{Z}_2)$. It shows how the remaining symmetry can be broken by Wilson lines, in a right-handed sneutrino direction, such that only the SM gauge group remains, with proto-realistic fermion mass matrices.
Final comments, lines of interest for future explorations, and the conclusions are presented in section~\ref{sec:conclusion}.

\section{$E_8$ SYM in 10d}
\label{sec:susy}
The theory of interest is $\mathcal{N}=1$ SYM theory in $10$d for the gauge group $E_8$.
It is assumed that all SM matter and gauge fields are unified into one $10$d vector gauge superfield $\mathcal{V}(x,z_1,z_2,z_3)$ (where $x$ denotes the uncompactified 4d coordinates in $R^4$ 
and the $z_i$ denote three complex coordinates of the remaining
compact $6$d space) that decomposes into a $10$d vector field and a $10$d Majorana fermion (which in $10$d is also a Weyl fermion). The basic hypothesis is that all SM matter, Higgs and gauge fields are unified into
a single ``particle'', namely the $10$d vector superfield: the $\mathcal{V}_{248}\sim  (\textbf{248})$ representation of $E_8$.
In $E_8$ the \textbf{248} is both the adjoint and the fundamental representation and it is real. The $10$d vector superfield $\mathcal{V}$ decomposes into a $4$d vector superfield $V$ and three $4$d chiral superfield multiplets $\phi_{1,2,3}$. 
In general, $\mathcal{N}=1$ SUSY in $n=7,8,9,10$ dimensions, implies $\mathcal{N}=4$ SUSY in 4d after compactification~\cite{ArkaniHamed:2001tb,Brink:1976bc}. In particular this implies that $E_8$ SYM theory with $\mathcal{N}=1$ SUSY in $10$d in principle has $\mathcal{N}=4$ SUSY in 4d after compactification to a compact 6d torus $T^6$. 
The $\mathcal{N}=4$ SYM Lagrangian in $R^4\times T^6$ is displayed first, followed by a description of how the extra supersymmetries may be broken by orbifolding.

\subsection{The $\mathcal{N}=4$ SYM Lagrangian in $R^4\times T^6$}

The extra dimensional space $T^6$ (assumed here to be a six dimensional torus) is parametrized with the three complex coordinates $z_{1,2,3}$ as $(T^2)^3$. These are called symmetric toroidal orbifolds. After compactification to the torus $T^6$, the $10$d real vector and $10$d Majorana fermion components of the supermultiplet decompose into one $4$d real vector $A_\mu$, six $4$d real scalars $X$ and four $4$d Weyl fermions $\lambda$. The compactified $\mathcal{N}=4$ SYM ($4$d) Lagrangian is
\begin{equation}\label{ec:N4SYM-lagrangian}
\begin{split}
\mathcal{L} &= -\frac{1}{2g^2}F_{\mu\nu}F^{\mu\nu}+\frac{\theta_I}{8\pi^2}F_{\mu\nu}\bar{F}^{\mu\nu}- i \overline{\lambda}^a\overline{\sigma}^\mu D_\mu \lambda_a -D_\mu X^i D^\mu X^i
\\ &\quad+g C^{ab}_i \lambda_a[X^i,\lambda_b] + g \overline{C}_{iab}\overline{\lambda}^a[X^i,\overline{\lambda}^b]+\frac{g^2}{2}[X^i,X^j]^2,
\end{split}\end{equation}
where $i,j=1...6$; $a,b=1...4$, and $C^{ab}_i$ represent the $SU(4)_R$ structure constants. The $SU(4)_R\simeq SO(6)_R$ symmetry comes as a remnant of the $6$d rotation group of the extra dimensions $O(6)$.
There are only two gauge coupling constants $g$ and $\theta_I$, and all the vertices are completely defined by them.

As the compactification will actually be in $(T^2)^3$, the theory can be more conveniently seen in terms of simple $\mathcal{N}=1$ SUSY by having one gauge vector supermultiplet $V$ and three chiral supermultiplets $\phi^i$:
\begin{equation}\begin{split}
\mathcal{L} &=\frac{1}{32}  \tau\int d^2\theta\  W^\alpha W_\alpha+\int d^2\theta d^2\bar{\theta}\bar{\phi}^i e^{2gV}\phi^i-\int d^2\theta\sqrt{2} g \phi_1[\phi_2,\phi_3]+h.c. ,
\label{eq:n41lag}
\end{split}\end{equation}
where $i= 1,2,3$. Note the explicit $SU(3)_R\times U(1)_R$ symmetry remaining from the rotation between the three complex coordinates and the complex rotation in all of them. This is particularly helpful since one can relate each chiral supermultiplet to the degrees of freedom of the vector superfield that come from each complex extra dimension.

\subsection{SUSY breaking by orbifolding}
\label{sec:orbo}

The extra dimensions are assumed to be orbifolded by a discrete group $F$ so that the actual extra dimensional space is $T^6/F$. In general, there are six extra dimensions with Poincar\`e symmetry $O(6)\ltimes T^6/\Gamma$, where $O(6)$ are rotations, and $T^6$ are the  translations. The translation group is modded by the lattice vectors $\Gamma=\mathbb{Z}^6$ which makes it compact $\mathbb{R}^6\to T^6\simeq\mathbb{R}^6/\Gamma$. Orbifolding means modding the rotation group by a discrete subgroup $F\in O(6)$. The group $F$ must be a symmetry of the lattice $F\Gamma=\Gamma$ to consistently define an orbifold. The rotation group is $O(6)\simeq SO(6)\times \mathbb{Z}_2\simeq SU(4)\times \mathbb{Z}_2$. 
 If one desires to keep some SUSY after orbifolding, then it can be done only by a discrete subgroup of $SU(3)$~\cite{Dixon:1985jw,Dixon:1986jc}. A simple $\mathbb{Z}_2$, for example, would break $\mathcal{N}=4\to \mathcal{N}=2 $ SUSY. 

The simplest way to break $\mathcal{N}=4\to \mathcal{N}=1 $ SUSY is with the $\mathbb{Z}_3$ orbifolding, 
\begin{equation}\begin{split}
(x,z_1,z_2,z_3)&\sim (x,\omega z_1,\omega z_2,\omega z_3),\\
\mathcal{V}(x,z_1,z_2,z_3)&= R(\omega)\mathcal{V}(x,\omega z_1,\omega z_2,\omega z_3),
\end{split}\end{equation}
where $\omega=e^{2i\pi/3}$ and $R(\omega)$ is the representation of the $\mathbb{Z}_3$ rotation acting on the vector superfield. In terms of the components of this superfield, the orbifolding acts as
\begin{equation}\begin{split}
V(x,z_1,z_2,z_3)&=V(x,\omega z_1,\omega z_2,\omega z_3),\\
\phi^i(x,z_1,z_2,z_3) &=\omega \phi^i(x,\omega z_1,\omega z_2,\omega z_3).
\end{split}\end{equation}
Note that, as desired, this orbifolding leaves invariant the Lagrangian in eq.~\eqref{eq:n41lag}. Each chiral multiplet is associated with the extra dimensional degrees of freedom of the vector super multiplet. Multiplying a complex coordinate must also multiply the corresponding components of a vector, which in this case are the chiral supermultiplets.
This orbifolding would only leave as a zero mode the $\mathcal{N}=1$ vector multiplet $V$, therefore providing a pure $\mathcal{N}=1$ SYM at low energies. In order to get the SM (or MSSM) one needs to consider more general orbifoldings than
$\mathbb{Z}_3$.

A more general orbifolding that preserves ``simple'' ($\mathcal{N}=1$) SUSY is 
\begin{equation}
F\simeq \mathbb{Z}_N \subset SU(3),
\end{equation}
with a positive integer $N$. 
A general $\mathbb{Z}_N$ orbifolding can be defined as
\begin{equation}\begin{split}
(x,z_1,z_2,z_3)&\sim (x,e^{2i\pi n_1/N} z_1, e^{2i\pi n_2/N} z_2, e^{2i\pi n_3/N} z_3)\\
\mathcal{V}(x,z_1,z_2,z_3)&= R(e^{2i\pi n_1/N},e^{2i\pi n_2/N},e^{2i\pi n_3/N} )\mathcal{V}(x,e^{2i\pi n_1/N} z_1,e^{2i\pi n_2/N} z_2,e^{2i\pi n_3/N} z_3),
\label{eq:Rorb}
\end{split}\end{equation}
where, as before, $R$ is the representation of the $\mathbb{Z}_N$ rotation acting on the $10$d vector superfield. For the transformation to belong to $SU(3)$ it must satisfy 
\begin{equation}
n_1+n_2+n_3=0\mod N,
\label{eq:nio}
\end{equation}
 so that it has determinant $1$. If we want SUSY to be preserved, the actual constraint would be
 \begin{equation}
n_1+n_2+n_3=0\mod 2N,
\label{eq:ni}
\end{equation}
 since fermions and scalars are rotated together, while fermions rotate twice as slow \cite{GrootNibbelink:2017luf}. Also note that can be $n_i> N$. 
 This orbifolding decomposes the fields as
\begin{equation}\begin{split}
V(x,z_1,z_2,z_3)&=V(x,e^{2i\pi n_1/N} z_1,e^{2i\pi n_2/N} z_2,e^{2i\pi n_3/N} z_3),\\
\phi^i(x,z_1,z_2,z_3) &=e^{2i\pi n_i/N} \phi^i(x,e^{2i\pi n_1/N} z_1,e^{2i\pi n_2/N} z_2,e^{2i\pi n_3/N} z_3).
\end{split}\end{equation}
By choosing the $n_i$, one chooses the chiral supermultiplets that do have zero modes and those that do not.

Finally note that the most general Abelian orbifolding that preserves $\mathcal{N}=1 $ SUSY is 
$F\simeq \mathbb{Z}_N \times  \mathbb{Z}_M\subset SU(3)$,
with positive integers $N,M$ (for discussion on the consistent Abelian and non Abelian orbifolds see \cite{deAnda:2019anb,Fischer:2012qj,Fischer:2013qza}). 
If preservation of $\mathcal{N}=1 $ SUSY is not required, then the most general Abelian group available for orbifolding would be $F\simeq\mathbb{Z}_N\times \mathbb{Z}_M\times \mathbb{Z}_L\times \mathbb{Z}_2$. 
The additional $\mathbb{Z}_L$ belongs to $U(3)$, with general positive integer $L$, and it need not satisfy eq.~\eqref{eq:ni}. This orbifolding breaks SUSY completely by adding a phase (the non identity determinant of the corresponding operation) to the $\theta$ coordinate and thus killing all fermionic zero modes.
The extra $\mathbb{Z}_2$ is related to reflection in the extra dimensions, and in this notation would involve complex conjugation of the $z_i$. Thus $\mathcal{N}=1 $ SUSY is broken while preserving an R parity identified as $\mathbb{Z}_2$.
This could be used to stabilize a dark matter candidate, for example.

\section{$\mathbf{E_8}$ breaking}
\label{sec:e8}

This section contains the proposal and discussion of the general orbifold $T^6/(\mathbb{Z}_N\times \mathbb{Z}_M)$ and how 
it can break the gauge group ${E_8}$ in various directions, while preserving $\mathcal{N}=1 $ SUSY.
To recap, the starting point is $\mathcal{N}=1$ SYM with a single $10$d vector superfield in the adjoint representation: 
$\mathcal{V}_{\textbf{248}}\sim  (\textbf{248})$.
In $E_8$ this is also the fundamental representation and it is real. 
As previously seen, assuming that $\mathcal{N}=1$ SUSY is preserved in the 4d theory, 
the $10$d vector superfield $\mathcal{V}$ decomposes into a $4$d vector superfield $V$ and three $4$d chiral superfield multiplets $\phi_{1,2,3}$,
\begin{equation}
\mathcal{V}_{\textbf{248}} \rightarrow \{V,\phi_{1,2,3}\}_{\textbf{248}},
\label{vector}
\end{equation}
where the chiral scalar superfields $\phi_{1,2,3}$ arise from the extra dimensional components of the original
10d gauge field $\mathcal{V}$. The task now is to also break the ${E_8}$ gauge theory into different subgroups,
such that the $\textbf{248}$ splinters into representations of the smaller subgroup, with some components surviving as zero
(massless) modes and other components only having (very) heavy massive modes, making them practically unobservable,
allowing suitable for
applications to particle physics with the massless modes identified as the starting point for various 4d models.

The model requires the extra dimensions to form the orbifold $T^6/F$, where $F$ is a discrete subgroup of $F\subset SU(4)\times \mathbb{Z}_2$. Assuming the orbifold to be Abelian and to preserve $\mathcal{N}=1 $ SUSY, one is led to consider the general orbifold,
\begin{equation}
F\simeq \mathbb{Z}_N\times \mathbb{Z}_M \subset SU(3),
\end{equation}
for positive integers $N$ and $M$, as defined in the next subsection.
However, the analysis is performed by  first considering a single $T^6/\mathbb{Z}_N$ orbifolding in the next subsection.

\subsection{$\mathbf{E_8}$ breaking by $T^6/\mathbb{Z}_N$ orbifolding}
\label{sec:orbt6e8}

The standard mechanism to break a gauge symmetry through a $\mathbb{Z}_N$ orbifolding is by adding a gauge transformation to the orbifold transformation. Since $\mathbb{Z}_N\subset U(1)$, one can assume that $\mathbb{Z}_N$ is accompanied by a specific $U(1)_a\subset E_8$ transformation. This would break the original gauge symmetry into the subgroup that commutes with $U(1)_a$.
 Let $q^f_a$ be the charge of a field $f$ under the chosen $U(1)_a$. Applying the boundary condition
\begin{equation}
f\sim e^{2i \pi q^f_a/N} f,
\end{equation}
breaks the symmetry consistently into a subgroup preserving the $U(1)_a$ (e.g. the multiplicative phase would correspond to a simple parity of $\pm 1$ for the simplest examples based on an orbifold parity of $\mathbb{Z}_2$).

A general $\mathbb{Z}_N$ orbifolding is defined by identifying
\begin{equation}
(x,z_1,z_2,z_3)\sim (x,e^{2i \pi n_1/N} z_1,e^{2i \pi n_2/N}z_2,e^{2i \pi n_3/N}z_3),
\label{ni}
\end{equation}
with arbitrary integers $n_i$ satisfying $n_1+n_2+n_3=0\ {\rm mod}\ 2N$. The decomposed $10$d superfield is transformed as
\begin{equation}\begin{split}
V(x,z_1,z_2,z_3)&=e^{2i \pi q^f_a/N} V(x,e^{2i\pi n_1/N} z_1,e^{2i\pi n_2/N} z_2,e^{2i\pi n_3/N} z_3),\\
\phi^i(x,z_1,z_2,z_3) &=e^{2i\pi n_i/N}e^{2i \pi q^f_a/N} \phi^i(x,e^{2i\pi n_1/N} z_1,e^{2i\pi n_2/N} z_2,e^{2i\pi n_3/N} z_3),
\end{split}
\label{phi^i}
\end{equation}
so that each multiplet is multiplied by a phase associated to the multiplet itself and to its charge. The representation of  $V$ is that of the adjoint of the unbroken gauge group, i.e. the fields with identity boundary conditions. The representation of the light chiral superfield $\phi_i$ is that of the fields with charge $q^f_a=-n_i \mod N$. One then chooses the $n_i$ to leave the desired light fields. 
This defines the orbifolding.

$E_8$ has rank $8$ and the orbifolding must preserve the SM gauge symmetry, $SU(3)_C\times SU(2)_L\times U(1)_Y$, which has rank $4$. This means that there are four different $U(1)$ groups (in addition to $U(1)_{Y}$, of course) that commute with the SM. One may define them by following the exceptional sequence~\cite{Slansky:1981yr}
\begin{equation}
\begin{split}
E_8&\supset E_7\times U(1)_F\\
&\quad \supset E_6\times U(1)_{F'}\times U(1)_F\\
&\quad\supset  SO(10)\times U(1)_{X'}\times U(1)_{F'}\times U(1)_F\\
&\quad\supset SU(5)\times U(1)_X\times U(1)_{X'}\times U(1)_{F'}\times U(1)_F\\
&\quad \supset SU(3)_C\times SU(2)_L\times U(1)_Y\times U(1)_X\times U(1)_{X'}\times U(1)_{F'}\times U(1)_{F},
\end{split}
\end{equation}
where any of the  (or a linear combination) can be chosen to be the $\mathbb{Z}_N$ orbifold operation
\begin{equation}
 \mathbb{Z}_N\subset U(1)_{Y}\times U(1)_X\times U(1)_{X'}\times U(1)_{F'}\times U(1)_F.
\end{equation}
 Depending on which one is chosen, and the order $N$, one obtains a certain preserved group. This is shown in detail in Table~\ref{tab:e8} where the breakings are presented for all the different choices from $\mathbb{Z}_2$ to $\mathbb{Z}_7$ (any orbifolding with a group larger than $\mathbb{Z}_7$ would not break further the symmetry).

\begin{table}[ht]
	\centering {\tiny
	\begin{tabular}[t]{l|ccccc}
		\hline
		 & $U(1)_Y$ & $U(1)_X$ & $U(1)_{X'}$  & $U(1)_{F'}$ & $U(1)_F$\\ 
		\hline
	$\mathbb{Z}_2 $ & $E_7\times SU(2)$ & $SO(16)$ & $SO(16)$   & $E_7\times SU(2)$& $E_7\times SU(2)$\\
	$\mathbb{Z}_3 $ & $E_6\times SU(3)$ & $SU(9)$  & $E_6\times SU(3)$ & $E_6\times SU(3)$ & $E_7\times U(1)$\\
	$\mathbb{Z}_4 $ &$SU(8)\times SU(2)$ & $SO(10)\times SU(4)$ & $SO(10)\times SU(4)$  &$E_6\times SU(2)\times U(1)$ &$E_7\times U(1)$\\
	$\mathbb{Z}_5 $ &$ SU(5)\times SU(5)$ &$ SU(5)\times SU(5)$& $SO(10)\times SU(3)\times U(1)$  & $E_6\times SU(2)\times U(1)$ &$E_7\times U(1)$\\
	$\mathbb{Z}_6 $ & $ SU(6)\times SU(3)\times SU(2)$& $SU(5)\times SU(4)\times U(1)$& $SO(10)\times SU(3)\times U(1)$  & $E_6\times SU(2)\times U(1)$ & $E_7\times U(1)$ \\
		$\mathbb{Z}_7 $ & $ SU(5)\times SU(3)\times SU(2)\times U(1)$& $SU(5)\times SU(4)\times U(1)$& $SO(10)\times SU(3)\times U(1)$  & $E_6\times SU(2)\times U(1)$& $E_7\times U(1)$ \\
			\hline
	\end{tabular}}
	\caption{$E_8$ breaking by orbifolding with $\mathbb{Z}_N\in U(1)_a$, where $a=Y,X,X',F',F$. The corresponding $\mathbb{Z}_N$ is a symmetry of the resulting group. Note there are no consistent orbifoldings for $\mathbb{Z}_{5}$ \cite{Fischer:2012qj}.} 
	\label{tab:e8}
\end{table}

By considering the orbifold $T^6/(\mathbb{Z}_N\times \mathbb{Z}_M)$, with combinations of $N,M$ selected from 
Table~\ref{tab:e8}, various patterns of symmetry breaking can be achieved consistently while preserving an
$\mathcal{N}=1 $ SUSY.
The idea is that the preserved group is the intersection of the preserved groups shown for individual values of 
$N,M$ chosen from Table~\ref{tab:e8}. 
However achieving the desired symmetry breaking pattern is not enough: it is also necessary to obtain the required chiral matter multiplets, and this can only be determined case by case.
The strategy consists of searching specific numbers that define the orbifold that breaks the ${\cal N}=4$ SUSY $E_8$ into a smaller group. Particular examples are discussed later.

\subsection{$\mathbf{E_8}$ breaking by a more general $T^6/(\mathbb{Z}_N\times \mathbb{Z}_M)$ orbifolding  }

It is clear that the SM field content is contained in the single 
$10$d vector superfield $\mathcal{V}_{248}\sim  (\textbf{248})$,
however there are many more fields present. The SM  has an $SO(3,1)$ Lorentz symmetry and $SU(3)_C\times SU(2)_L\times U(1)_Y$ gauge symmetry. Under these symmetries, the SM field content is made of $19$ multiplets ($g_\mu, W_\mu,B_\mu,Q_i,L_i,e^c_i,d^c_i, u^c_i, H$).

A viable model should resemble the SM, with its field content and their couplings, at low energies. Since the initial symmetry is strongly restrictive, it is up to the structure of the orbifold to weed out the extra fields and generate the effective couplings after compactification.

Consider a more general orbifold $T^6/(\mathbb{Z}_N\times \mathbb{Z}_M)$, where 
\footnote{Note such orbifolds are also constrained by lattice consistency~\cite{Fischer:2012qj}, which we check case by case.}
\begin{equation}
\mathbb{Z}_N\times \mathbb{Z}_M\subset U(1)_{Y}\times U(1)_X\times U(1)_{X'}\times U(1)_{F'}\times U(1)_{F}.
\end{equation}
This orbifold is defined by linear combinations of charges
\begin{equation}
\mathbb{Z}_N: \phi\to e^{2i\pi (a q_Y+b q_X+c q_{X'}+d q_{F'}+e q_F)/N}\phi,\ \ \   \mathbb{Z}_M: \phi\to e^{2i\pi  (g q_Y+hq_X+jq_{X'}+kq_{F'}+lq_F)/M}\phi,
\end{equation}
which are applied as
\begin{equation}\begin{split}
(x,z_1,z_2,z_3)&\sim (x,e^{2in_1 \pi/N} z_1,e^{2in_2 \pi/N} z_2,e^{2in_3 \pi/N} z_3),\\
(x,z_1,z_2,z_3)&\sim (x,e^{2im_1 \pi/M}z_1,e^{2im_2 \pi/M} z_2,e^{2im_3 \pi/M} z_3),
\end{split}\end{equation}
with arbitrary positive integers $N$ and $M$, arbitrary integers $(a,b,c,d,e,g,h,j,k,l)$, and $n_1+n_2+n_3=m_1+m_2+m_3=0\ {\rm mod} \ 2N$, preserving SUSY. 

In the next subsection, it is shown that, although the orbifolding by itself can never break to the SM gauge group directly, nevertheless
this can in principle be achieved by Wilson lines.

\subsection{Wilson Lines and Scalar Vacuum Expectation Values}

It is apparent that 
no orbifolding can ever break $E_8\to SU(3)_C\times SU(2)_L\times U(1)_Y$ directly. 
The reason is clear: orbifolding by itself cannot reduce the rank of the group \cite{Forste:2005rs,Hebecker:2003jt,Aranda:2019nac}.
However, it is possible to add non trivial gauge transformations to whole loop translation in each circle. Adding these phases creates Wilson lines that generate effective vaccum expectation values (VEVs) for the scalars coming from the extra dimensional part of gauge fields, dynamically breaking the symmetry~\cite{Candelas:1985en}. 

According to the Wilson line mechanism, one can also reduce break the symmetry by adding a gauge transformation to the extra dimensional translations. The extra dimensional torus is defined as
\begin{equation}
T^6\simeq \mathbb{R}^6/\Gamma,\ \ \ {\rm where}\ \ \ \Gamma=\{\gamma_i^r\},
\end{equation}
where $i=1,2,3$ and $r=1,2$. The $\gamma_i^r$ are the basis vectors, separated into three pairs for the three torii. They define the torii periodicity as
\begin{equation}
z_i=z_i+\gamma_i^r.
\end{equation}
The 10 gauge superfield must also comply with the periodicity, up to a gauge transformation $U$
\begin{equation}
\mathcal{V}(x,z_i)=U^r_i \mathcal{V}(x,z_i+\gamma^r_i).
\label{eq:vwl}
\end{equation}
The choice of a non trivial $U^r_i$ gauge transformation is the Wilson line mechanism.

In a factorisable orbifold, as assumed in this paper,
there are six different circles and six independent phases $U_i^r$ with defined by
\begin{equation}
U_i^r=e^{i\alpha^{ar}_{i} \gamma^r_i  T_a},
\end{equation}
where $T_a$ are the generators of the gauge group and $a$ runs through its orded. No sum intended on $i,r$.
These phases, being associated with translations, must commute with each other
\begin{equation}
[U_i^r,U_j^s]=0.
\end{equation}
The eq. \ref{eq:vwl} fixes the superfield to have the dependence on the extra dimensions
\begin{equation}
\mathcal{V}(x,z_i)=e^{i\alpha^{ar}_i z_i  T_a}\tilde{\mathcal{V}}(x,z_i), \ \ \ {\rm with}\ \ \ \tilde{\mathcal{V}}(x,z_i)=\tilde{\mathcal{V}}(x,z_i+\gamma^r_i).
\end{equation}
The Wilson lines can be reabsorbed through a gauge transformation $\mathcal{V}\to U^{-1}\mathcal{V}$ generating VEVs in the extra dimensional components of the gauge vectors. Since preserving SUSY is desired in this case, the six extra-dimensional components of the gauge vector arrange themselves into the three complex scalar components of the resulting three chiral supermultiplets.
After the gauge transformation absorbing the phase, the components become
 components
\begin{equation}
\begin{split}
V(x,z_i)&=\tilde{V}(x,z_i),\\
\phi_i(x,z_i)&=\tilde{\phi}(x,z_i)+\sum_r \alpha^{ar}_{i} \gamma^r_i  T_a,
\end{split}
\end{equation}
therefore the chiral supermultiplets have now a VEV
\begin{equation}
\braket{\phi_i^a}=\alpha^a_i,
\end{equation}
with $\alpha^a_i=\sum_r \alpha^{ar}_{i} \gamma^r_i$ a complex number: Since it involves the torii basis vectors, this VEV is associated with the torii radii and therefore its natural scale is the compactification scale. Therefore the Wilson line mechanism induces Spontaneous Symmetry Breaking at the compactification scale, just as a high scale Higgs mechanism would.

The VEV comes from the phase $U^r_i$ related to translations. Note that for consistency with the orbifold (i.e. they must fulfill the rotation-translation commutation relations coming from the Poincar\`e algebra) boundary conditions, the VEVs must lie in chiral supermultiplets with a zero mode: the VEVs must be aligned in the gauge representations that have a zero mode. By integrating out the other fields, one can obtain an effective potential for the fields that get a VEV~
\cite{Hosotani:1983xw,Hosotani:1983vn,Hosotani:2004wv,Hosotani:2004ka,Haba:2004qf,Haba:2002py}. The $D=0$ and $F=0$ flatness 
conditions must also be satisfied by the VEVs if SUSY is to be preserved.

In conclusion, using the Wilson line mechanism one can give VEVs for the chiral supermultiplets with zero modes,
and perhaps even the massive modes, through three independent commuting phases 
$\alpha^a_i$. Such continuous Wilson lines, not in the direction of the adjoint representation of the remaining gauge group, reduce the rank. Since orbifolding does not reduce rank, the symmetry breaking induced by the continuous Wilson lines is a crucial ingredient of the model.~\footnote{Note that discrete Wilson lines (not considered here) would also not reduce the rank, even though they could break the gauge symmetry.}

\section{Examples of $E_8$ breaking}
\label{examples}

The biggest challenge when building a unification model based on $E_8$ is the fact that it only has real representations. This is a problem because fermions in the SM are chiral, and so far there is no evidence for the existence of mirror fermions.
Orbifolding gives a way to overcome this problem. The simplest model found, that obtains a chiral representation from $E_8$,  ends up with $E_6$ as the remaining gauge symmetry after compactification.  To see how this happens, it is convenient to analyze how the representations decompose into the different subgroup representation due to a particular orbifold. This gives the information needed to determine what fields survive in the low energy theory.
This section contains a series of examples to illustrate the procedure one must follow to obtain a realistic model from 
the preceding formalism in the previous section,
starting with the simplest $E_6$ example, before moving on to other examples.
These examples are not meant to be realistic, for example they do not have the desired matter content of the SM and 
are not anomaly free.
In the next section we will discuss a realistic example.

\subsection{$E_6\times SU(3)_{F}$ from $T^6/\mathbb{Z}_3$}
\label{4.1}
An interesting possibility is the orbifold 
$T^6/\mathbb{Z}_3$~\cite{Babu:2002ti,Ahsan:2010zt,Kobayashi:2017fgl},
where the $\mathbb{Z}_3 \subset U(1)_{X'}$
orbifolding from Table~\ref{tab:e8}
breaks $E_8\to E_6\times SU(3)_{F}$, where 
\begin{equation}
\textbf{248} \to (\textbf{78},\textbf{1})+(\textbf{1},\textbf{8})+(\textbf{27},\textbf{3})+(\bar{\textbf{27}},\bar{\textbf{3}}).
\end{equation}
The $U(1)_{X'}$ charges are identified by considering the hypothetical 
(but unachievable as shown below) further decomposition into $SO(10)\times U(1)_{X'}\times SU(3)_F$ representations as
\begin{equation}\begin{split}
(\textbf{78},\textbf{1}) &\to(\textbf{45},0,\textbf{1})+(\textbf{1},0,\textbf{1})+ (\textbf{16},-3,\textbf{1})+(\bar{\textbf{16}},3,\textbf{1}),\\
(\textbf{1},\textbf{8}) &\to(\textbf{1},0,\textbf{8}),\\
(\textbf{27},\textbf{3}) &\to (\textbf{16},1,\textbf{3})+(\textbf{10},-2,\textbf{3})+(\textbf{1},4,\textbf{3}),\\
(\bar{\textbf{27}},\bar{\textbf{3}}) &\to (\bar{\textbf{16}},-1,\bar{\textbf{3}})+(\textbf{10},2,\bar{\textbf{3}})+(\textbf{1},-4,\bar{\textbf{3}}),
\end{split}\end{equation}
where one can see that each
$E_6\times SU(3)_{F}$ representation has the same $U(1)_{X'}$ charge mod $3$, 
corresponding to the $\mathbb{Z}_3$
symmetry. 
For example, $(\textbf{78},\textbf{1})$ and $(\textbf{1},\textbf{8})$ have $U(1)_{X'}$ charge zero mod $3$, corresponding to 
a $\mathbb{Z}_3$ singlet (denoted as $1$).
Therefore this orbifold indeed breaks $E_8\to E_6\times SU(3)_{F}$ and not $E_8\to
SO(10)\times U(1)_{X'}\times SU(3)_F$.

Under $E_8\to E_6\times SU(3)_{F}$, the single $10$d vector superfield in the adjoint representation
$\mathcal{V}_{\textbf{248}}\sim  \textbf{248}$
is separated into $E_6\times SU(3)_{F}$ multiplets with three different $\mathbb{Z}_3$ charges, listed in each line
\begin{equation}\begin{split}
\mathcal{V}_{\textbf{248}}
 &\to  \mathcal{V}_{(\textbf{78},\textbf{1})}+ \mathcal{V}_{(\textbf{1},\textbf{8})}\\
 &\quad+  \mathcal{V}_{(\textbf{27},\textbf{3})}\\
 &\quad+  \mathcal{V}_{(\bar{\textbf{27}},\bar{\textbf{3}})},
\end{split}\end{equation}
where $ \mathcal{V}_{(\textbf{78},\textbf{1})}$ and $\mathcal{V}_{(\textbf{1},\textbf{8})}$ are $\mathbb{Z}_3$ charge $1$ (singlets),  while $\mathcal{V}_{(\textbf{27},\textbf{3})}$ has one unit of $\mathbb{Z}_3$ charge 
$\omega=e^{2i\pi/3}$, and $ \mathcal{V}_{(\bar{\textbf{27}},\bar{\textbf{3}})}$ has $\mathbb{Z}_3$
charge $\omega^2$.

The action of the orbifold is defined as
\begin{equation}
(x,z_1,z_2,z_3)\sim (x,\omega^2 z_1,\omega^2 z_2,\omega^2 z_3),
\end{equation}
where $\omega=e^{2i\pi/3}$
as well as 
\begin{equation}
\mathbb{Z}_3: \mathcal{V}\to e^{2i\pi q_{X'}/3}\mathcal{V}.
\end{equation}
The orbifold decomposes the $10$d vector superfield to $\mathcal{N}=1$ vector and chiral superfields,
\begin{equation}
\mathcal{V} \rightarrow \{V,\phi_{1,2,3}\},
\end{equation}
where the chiral superfields $\phi_{1,2,3}$ are associated with the extra dimensions $z_1,z_2,z_3$,
respectively.
The resultant charges, under the combined action of the orbifold and $\mathbb{Z}_3$,
for each $\mathcal{N}=1$ 
multiplet are listed in table~\ref{tab:pe}. This is a simple application of the general formula in eq.~\eqref{phi^i}.
Therefore after compactification, at low energies, one has 
$\mathcal{N}=1$ SUSY with $E_6\times SU(3)_F$, where the final - low energy - field content is composed of the fields in table~\ref{tab:pe} with charge $1$. Note that there are fermions with zero modes
contained in ${\phi_i}_{ (\textbf{27},\textbf{3})}$.
Note also that, importantly, there are no mirror fermions with zero modes and one then obtains chiral matter from a real representation~\cite{Fonseca:2015aoa}.
\begin{table}[ht]
	\centering
	\begin{tabular}[t]{l|cccc}
		\hline
		 & $V$ & $\phi_1$ & $\phi_2$ & $\phi_3$\\ 
		\hline
	$\mathcal{V}_{(\textbf{78},\textbf{1})} $ & $1$ & $\omega^2$ & $\omega^2$& $\omega^2$\\
		$\mathcal{V}_{(\textbf{1},\textbf{8})} $ & $1$ & $\omega^2$ & $\omega^2$& $\omega^2$\\
	$\mathcal{V}_{(\textbf{27},\textbf{3})} $ & $\omega$ & $1$ & $1$ & $1$\\
	$\mathcal{V}_{(\bar{\textbf{27}},\bar{\textbf{3}})} $ & $\omega^2$ & $\omega$& $\omega$& $\omega$ \\
		\hline
	\end{tabular}
	\caption{Charges of each $\mathcal{N}=1$ superfield $E_6\times SU(3)_{F}$ multiplet under the 
	$\mathbb{Z}_3$ orbifolding . 
	Only the charge singlets (denoted by charges $1$) have massless zero
	modes and survive in the low energy theory.} 
	\label{tab:pe}
\end{table}
\begin{table}
	\centering
	\footnotesize
	\renewcommand{\arraystretch}{1.1}
	\begin{tabular}[t]{l|llll}
		\hline
		 & $V$ & $\phi_1$ & $\phi_2$ & $\phi_3$\\ 
		\hline
	$\mathcal{V}_{(\textbf{78},0,0)} $ & $1,1$ & $\omega^2, 1$ & $\omega^2, \omega$& $\omega^2, \omega^2$\\
		$\mathcal{V}_{(\textbf{1},0,0)} $ & $1, 1$ & $\omega^2, 1$ & $\omega^2, \omega$& $\omega^2, \omega^2$\\
		$\mathcal{V}_{(\textbf{1},0,0)} $ & $1, 1$ & $\omega^2, 1$ & $\omega^2, \omega$& $\omega^2, \omega^2$\\
		$\mathcal{V}_{(\textbf{1},3,1)} $ & $1, \omega$& $\omega^2, \omega$ & $\omega^2, \omega^2$& $\omega^2, 1$\\
		$\mathcal{V}_{(\textbf{1},3,-1)} $ & $1, \omega^2$ & $\omega^2, \omega^2$ & $\omega^2, 1$& $\omega^2, \omega$\\
		$\mathcal{V}_{(\textbf{1},0,-2)} $& $1, \omega$& $\omega^2, \omega$ & $\omega^2, \omega^2$& $\omega^2, 1$\\
		$\mathcal{V}_{(\textbf{1},-3,-1)} $  & $1, \omega^2$ & $\omega^2, \omega^2$ & $\omega^2, 1$& $\omega^2, \omega$\\
		$\mathcal{V}_{(\textbf{1},-3,1)} $ & $1, \omega$& $\omega^2, \omega$ & $\omega^2, \omega^2$& $\omega^2, 1$\\
		$\mathcal{V}_{(\textbf{1},0,2)} $  & $1, \omega^2$ & $\omega^2, \omega^2$ & $\omega^2, 1$& $\omega^2, \omega$\\	
		\hline
	\end{tabular}
	\hspace*{0.5cm}
	\begin{tabular}[t]{l|llll}
		\hline
		 & $V$ & $\phi_1$ & $\phi_2$ & $\phi_3$\\ 
		\hline
	$\mathcal{V}_{(\textbf{27},1,1)} $ & $\omega, \omega^2$ & $1, \omega^2$ & $1, 1$ & $1, \omega$\\
	$\mathcal{V}_{(\textbf{27},1,-1)} $ & $\omega, 1$ & $1, 1$ & $1, \omega$ & $1, \omega^2$\\
	$\mathcal{V}_{(\textbf{27},-2,0)} $ & $\omega, \omega$ & $1, \omega$ & $1, \omega^2$ & $1, 1$\\
	$\mathcal{V}_{(\bar{\textbf{27}},-1,-1)} $ & $\omega^2, \omega$ & $\omega, \omega$& $\omega, \omega^2$& $\omega, 1$ \\
		$\mathcal{V}_{(\bar{\textbf{27}},-1,1)} $ & $\omega^2, 1$ & $\omega, 1$& $\omega, \omega$& $\omega, \omega^2$ \\
			$\mathcal{V}_{(\bar{\textbf{27}},2,0)} $ & $\omega^2, \omega^2$ & $\omega, \omega^2 $& $\omega, 1$& $\omega, \omega$ \\
		\hline
	\end{tabular}
	\caption{Charges of each $\mathcal{N}=1$ superfield $E_6\times U(1)_{F'}\times U(1)_F$ multiplet 
	under a $\mathbb{Z}_3\times \mathbb{Z}'_3$ orbifolding. Only the fields with both charges 
	equal to unity have zero modes.} 
	\label{tab:pe2}
\end{table}

Thus, orbifolding has produced a chiral representation from a real one. This is good news, however it is not quite what one needs yet. As table~\ref{tab:pe} shows, the orbifolding leaves three copies of triplets, ${\phi_i}_{ (\textbf{27},\textbf{3})}$ with $i=1,2,3$, 
therefore providing nine light families, which are too many.

\subsection{$E_6$ based model from $T^6/(\mathbb{Z}_3\times \mathbb{Z}'_3)$}
\label{sec:e6ssm}
A further problem of the previous example, is that the family symmetry $SU(3)_F$ is unbroken.
Starting from the setup in the previous section, another orbifolding $\mathbb{Z}_3$ is desired that breaks $SU(3)_F\to U(1)_{F'}\times U(1)_F$ with the representation decomposition
\begin{equation}\begin{split}
\textbf{3}&\to(1,1)+(1,-1)+(-2,0),\\
\textbf{8}&\to (0,0)+(0,0)+(3,1)+(3,-1)+(0,-2)+(-3,-1)+(-3,1)+(0,2).
\end{split}\end{equation}
Therefore we are led to consider $T^6/(\mathbb{Z}_3\times \mathbb{Z}'_3)$, where the first $\mathbb{Z}_3$ orbifolding is as in the previous subsection
 and the extra $\mathbb{Z}'_3$ orbifold operation is defined by
\begin{equation}\begin{split}
(x,z_1,z_2,z_3)&\sim (x, \omega^3z_1,\omega z_2,\omega^2 z_3),
\end{split}\end{equation}
with
\begin{equation}
\mathbb{Z}'_3: \phi\to e^{i\pi (q_F+q_{F'})/3}\phi.
\end{equation}
Note that the coordinate $z_1$ is rotated completely by $\omega^3=1$ to comply with eq. \ref{eq:ni}.
Following the same procedure as discussed in subsection~\ref{4.1}, this 
leaves only three light chiral superfields
as shown in table~\ref{tab:pe2}: 
${\phi_1}_{ (\textbf{27},1,-1)}$, ${\phi_2}_{ (\textbf{27},1,1)}$, ${\phi_3}_{ (\textbf{27},-2,0)}$.

After compactification, each $\textbf{27}$, contains a full \textbf{16} component family of fermions plus a \textbf{10} component multiplet 
which contains one pair of Higgs doublets per family, plus exotics, plus a singlet  \textbf{1}.
The resulting field theory after compactification, (although it does not preserve SUSY~\cite{Fischer:2012qj}), resembles the E$_6$SSM~\cite{King:2005jy} in its minimal version~\cite{Howl:2007zi}
(for a recent review see \cite{King:2020ldn}).  
Thus with $T^6/(\mathbb{Z}_3\times \mathbb{Z}'_3)$ the full SM (plus additional states of an $E_6$ GUT)
can result from the single $10$d $\textbf{248}$ vector superfield.

 \subsection{$SO(10)$ based model from $T^6/\mathbb{Z}_6$}
 \label{SO(10)}
We now consider using $\mathbb{Z}_6\in U(1)_{X'}$ for the orbifold. The
$\textbf{248}$ representation decomposes under $SO(10) \times SU(3)_{F} \times U(1)_{X'}$ as 
\begin{equation}\begin{split}
(\textbf{248}) &\to (\textbf{45},\textbf{1},0)+(\textbf{1},\textbf{1},0)+(\textbf{1},\textbf{8},0)\\
&\quad + (\textbf{16},\textbf{3},1)\\
&\quad +(\textbf{1},\bar{\textbf{3}},-4)+(\textbf{10},\bar{\textbf{3}},2)\\
&\quad + (\textbf{16},\textbf{1},-3)+(\bar{\textbf{16}},\textbf{1},3) \\
&\quad+(\textbf{1},\textbf{3},4)+(\textbf{10},\textbf{3},-2)\\
&\quad+ (\bar{\textbf{16}},\bar{\textbf{3}},-1).
 \label{eq:so10rep}
\end{split}\end{equation}
The fields in each of the six lines of eq. \ref{eq:so10rep}, have a common $\mathbb{Z}_6$ charge:
$1,\alpha, \alpha^2, \alpha^3, \alpha^4, \alpha^5$, respectively, where $\alpha=e^{2i\pi/6}$.
The orbifolding can be applied as
\begin{equation}\begin{split}
(x,z_1,z_2,z_3)&\sim (x, \alpha^5 z_1,\alpha^2 z_2,\alpha^5 z_3),
\end{split}\end{equation}
where $\alpha=e^{2i\pi/6}$ and
\begin{equation}
\mathbb{Z}_6: \phi\to e^{2i\pi q_{X'}/6}\phi.
\end{equation}
Following the same procedure as discussed in subsection~\ref{4.1},
this gives the charge of each $\mathcal{N}=1$ superfield as shown in table~\ref{tab:pe3}. 
\begin{table}
	\centering
	\footnotesize
	\renewcommand{\arraystretch}{1.1}
	\begin{tabular}[t]{l|cccc}
		\hline
		 & $V$ & $\phi_1$ & $\phi_2$ & $\phi_3$\\ 
		\hline
	$\mathcal{V}_{(\textbf{45},\textbf{1},0)} $ & $1$ & $\alpha^5$ & $\alpha^2$& $\alpha^5$\\
		$\mathcal{V}_{(\textbf{1},\textbf{1},0)} $ & $1$ & $\alpha^5$ & $\alpha^2$& $\alpha^5$\\
		$\mathcal{V}_{(\textbf{1},\textbf{8},0)} $ & $1$ & $\alpha^5$ & $\alpha^2$& $\alpha^5$\\
		$\mathcal{V}_{ (\textbf{16},\textbf{3},1)} $ & $\alpha$ & $1$ & $\alpha^3$& $1$\\
		$\mathcal{V}_{(\textbf{1},\bar{\textbf{3}},-4)} $ & $\alpha^2$ & $\alpha$ & $\alpha^4$& $\alpha$\\
		$\mathcal{V}_{(\textbf{10},\bar{\textbf{3}},2)} $& $\alpha^2$ & $\alpha$ & $\alpha^4$& $\alpha$\\
		\hline
	\end{tabular}
	\hspace*{0.5cm}
	\begin{tabular}[t]{l|cccc}
		\hline
		 & $V$ & $\phi_1$ & $\phi_2$ & $\phi_3$\\ 
		\hline
	$\mathcal{V}_{(\textbf{16},\textbf{1},-3)} $ & $\alpha^3$ & $\alpha^2$ & $\alpha^5$& $\alpha^2$\\
	$\mathcal{V}_{(\bar{\textbf{16}},\textbf{1},3) } $ & $\alpha^3$ & $\alpha^2$ & $\alpha^5$& $\alpha^2$\\
	$\mathcal{V}_{(\textbf{1},\textbf{3},4)} $ & $\alpha^4$ & $\alpha^3$ & $1$& $\alpha^3$\\
	$\mathcal{V}_{(\textbf{10},\textbf{3},-2)} $ & $\alpha^4$ & $\alpha^3$ & $1$& $\alpha^3$\\
		$\mathcal{V}_{(\bar{\textbf{16}},\bar{\textbf{3}},-1)} $ & $\alpha^5$ & $\alpha^4$ & $\alpha$& $\alpha^4$\\
		\hline
	\end{tabular}
	\caption{Charges of each $\mathcal{N}=1$ superfield $SO(10) \times SU(3)_{F} \times U(1)_{X'}$ multiplet 
	under a $\mathbb{Z}_6$ orbifolding. Only the fields with unit charges have zero modes, leading to an $SO(10)$ model after compactification.} 
	\label{tab:pe3}
\end{table}

Although ${\phi_2}_{(\textbf{10},\textbf{3},-2)}$ contains the MSSM Higgses,
both ${\phi_1}_{ (\textbf{16},\textbf{3},1)}$ and ${\phi_3}_{ (\textbf{16},\textbf{3},1)}$ have a zero mode
and each contain three SM fermion families. The model needs another orbifolding to further break the symmetry, 
and halve the fermion content. Although the other orbifolding cannot break the flavour symmetry as in sec. \ref{sec:e6ssm},
the model does contain a flavon ${\phi_2}_{(\textbf{1},\textbf{3},4)}$ which can do the job. The ingredients 
of the  above $T^6/\mathbb{Z}_6$ orbifold, when supplemented by a further $\mathbb{Z}_2$ orbifolding,
are the basis of the realistic model, described in sec. \ref{sec:psorb}.

 \subsection{$SU(5)$ based from $T^6/\mathbb{Z}_6$ }
\label{sec:zux}
One may also consider using $\mathbb{Z}_6\in U(1)_{X}$ for the orbifold.
The $\textbf{248}$ representation decomposes under $E_8\to SU(5)\times SU(4)\times U(1)_X$ as
 \begin{equation}\begin{split}
(\textbf{248}) &\to (\textbf{24},\textbf{1},0)+(\textbf{1},\textbf{1},0)+(\textbf{1},\textbf{15},0)\\
&\quad +(\textbf{1},\textbf{4},-5) +(\bar{\textbf{10}},\bar{\textbf{4}},1)\\
&\quad +(\bar{\textbf{10}},\textbf{1},-4)
+(\textbf{5},\textbf{6},2)\\
&\quad +(\bar{\textbf{5}},\textbf{4},3)+(\textbf{5},\bar{\textbf{4}},-3)\\
&\quad +(\textbf{10},\textbf{1},4)+(\bar{\textbf{5}},\textbf{6},-2)\\
&\quad +(\textbf{10},\textbf{4},-1)+(\textbf{1},\bar{\textbf{4}},5),
\end{split}\end{equation}
where the fields in the same line share the same $\mathbb{Z}_6$ charge: $1,\alpha, \alpha^2, \alpha^3, \alpha^4, \alpha^5$, respectively, where $\alpha=e^{2i\pi/6}$. One may now consider various choices of geometrical orbifoldings, leading to various massless modes. We have considered many such possibilities, but in the interests of brevity we do not display the results here.
Instead we highlight a common challenge to all such models.

Once a particular $SU(5)$ model is considered, one must consider the usual GUT problem of Doublet-Triplet splitting. Since there is only a single $\textbf{248}$ representation in the beginning, the field content and its symmetries are fixed. There are no representations to allow the Missing Partner mechanism~\cite{Masiero:1982fe}, nor the shaping symmetry to make the Dimopolous-Wilczeck mechanism\cite{Wilczeck:1982}. The only way to achieve it in this setup is through orbifolding. The next available $\mathbb{Z}_M$ orbifolding should break $SU(5)$ and achieve the Doublet-Splitting. However this leaves 
an unbroken $SU(4)$ flavour symmetry, and therefore any such setup generically 
contains 4 families at low energies.

\subsection{SM based model from $T^6/(\mathbb{Z}_N\times \mathbb{Z}_M)$}
\label{sec:smorb}

Finally, we discuss the results of a general scan of $T^6/(\mathbb{Z}_N\times \mathbb{Z}_M)$ orbifolds which 
can give rise to the SM gauge group factor, together with other gauge group factors.
We find that imposing the requirement that the field content with zero modes must have three families of fermions
does not allow the breaking of the $SU(3)_F$ family symmetry from orbifolding. Therefore,
without loss of generality, one can choose
linear combinations of charges restricted such that 
\begin{equation}
\mathbb{Z}_N\times \mathbb{Z}_M\subset U(1)_{Y}\times U(1)_X\times U(1)_{X'}.
\end{equation}
The decomposition under $E_8\to SU(3)_C\times SU(2)_L \times U(1)_Y\times U(1)_X\times U(1)_{X'}\times SU(3)_F$ 
of the $\textbf{248}$ representation is (using a colour coding explained below):
\footnote{Multiplying every flavour triplet by 3 and flavour adjoint by 8, one finds 99 SM multiplets and hence $1089$ SM states ($99\times (6\ {\rm scalars}+4\ {\rm fermions}+1\ {\rm vector}$)).}
\begin{equation}\begin{split}
(\textbf{248}) &\to\textcolor{magenta}{(\textbf{8},\textbf{1},0,0,0,\textbf{1})}+\textcolor{magenta}{(\textbf{1},\textbf{3},0,0,0,\textbf{1})}+\textcolor{magenta}{(\textbf{1},\textbf{1},0,0,0,\textbf{8})}\\
&\quad+\textcolor{magenta}{(\textbf{1},\textbf{1},0,0,0,\textbf{1})}+\textcolor{magenta}{(\textbf{1},\textbf{1},0,0,0,\textbf{1})}+\textcolor{magenta}{(\textbf{1},\textbf{1},0,0,0,\textbf{1})}\\
&\quad+(\textbf{1},\textbf{1},6,4,0,\textbf{1})+(\bar{\textbf{3}},\textbf{1},-4,4,0,\textbf{1})+(\textbf{3},\textbf{2},1,4,0,\textbf{1})\\
&\quad+(\textbf{1},\textbf{1},-6,-4,0,\textbf{1})+(\textbf{3},\textbf{1},4,-4,0,\textbf{1})+(\bar{\textbf{3}},\textbf{2},-1,-4,0,\textbf{1})\\
&\quad +(\textbf{1},\textbf{1},6,-1,-3,\textbf{1})+(\bar{\textbf{3}},\textbf{1},-4,-1,-3,\textbf{1})+(\textbf{3},\textbf{2},1,-1,-3,\textbf{1})\\
&\quad+(\textbf{1},\textbf{1},-6,1,3,\textbf{1})+(\textbf{3},\textbf{1},4,1,3,\textbf{1})+(\bar{\textbf{3}},\textbf{2},-1,1,3,\textbf{1})\\
&\quad+(\textbf{1},\textbf{2},-3,3,-3,\textbf{1})+(\bar{\textbf{3}},\textbf{1},2,3,-3,\textbf{1})+(\textbf{3},\textbf{2},-5,0,0,\textbf{1})\\
&\quad+(\textbf{1},\textbf{2},3,-3,3,\textbf{1})+(\textbf{3},\textbf{1},-2,-3,3,\textbf{1}) +(\bar{\textbf{3}},\textbf{2},5,0,0,\textbf{1})\\
&\quad+\textcolor{blue}{(\textbf{1},\textbf{1},6,-1,1,\textbf{3})}+\textcolor{blue}{(\bar{\textbf{3}},\textbf{1},-4,-1,1,\textbf{3})}+\textcolor{blue}{(\textbf{3},\textbf{2},1,-1,1,\textbf{3})}\\
&\quad+\textcolor{blue}{(\textbf{1},\textbf{2},-3,3,1,\textbf{3})}+\textcolor{blue}{(\bar{\textbf{3}},\textbf{1},2,3,1,\textbf{3})}+ \textcolor{violet}{(\textbf{1},\textbf{1},0,-5,1,\textbf{3})}\\
&\quad+\textcolor{red}{(\textbf{1},\textbf{1},-6,1,-1,\bar{\textbf{3}})}+\textcolor{red}{(\textbf{3},\textbf{1},4,1,-1,\bar{\textbf{3}})}+\textcolor{red}{(\bar{\textbf{3}},\textbf{2},-1,1,-1,\bar{\textbf{3}})}\\
&\quad+\textcolor{red}{(\textbf{1},\textbf{2},3,-3,-1,\bar{\textbf{3}})}+\textcolor{red}{(\textbf{3},\textbf{1},-2,-3,-1,\bar{\textbf{3}})}+ \textcolor{Dandelion}{(\textbf{1},\textbf{1},0,5,-1,\bar{\textbf{3}})}\\
&\quad +\textcolor{ForestGreen}{(\textbf{1},\textbf{2},3,2,-2,\textbf{3})}+(\textbf{3},\textbf{1},-2,2,-2,\textbf{3})+\textcolor{ForestGreen}{(\textbf{1},\textbf{2},-3,-2,-2,\textbf{3})}+(\bar{\textbf{3}},\textbf{1},2,-2,-2,\textbf{3})\\
&\quad+\textcolor{orange}{(\textbf{1},\textbf{2},-3,-2,2,\bar{\textbf{3}})}+(\bar{\textbf{3}},\textbf{1},2,-2,2,\bar{\textbf{3}})+\textcolor{orange}{(\textbf{1},\textbf{2},3,2,2,\bar{\textbf{3}})}+(\textbf{3},\textbf{1},-2,2,2,\bar{\textbf{3}})\\
&\quad+\textcolor{brown}{ (\textbf{1},\textbf{1},0,-5,-3,\textbf{1})}+\textcolor{brown}{(\textbf{1},\textbf{1},0,5,3,\textbf{1})}+\textcolor{brown}{(\textbf{1},\textbf{1},0,0,-4,\bar{\textbf{3}})}+\textcolor{brown}{(\textbf{1},\textbf{1},0,0,4,\textbf{3})}.
\label{eq:e8sm}
\end{split}\end{equation}
where the electric charge generator is given by $Q=T_{3L}+Y/6$ in our normalisation.

Thus one finds a unique (colour coded) embedding of the SM where:
\begin{itemize}
\item \textcolor{blue}{Blue} corresponds to the SM fermions and one family triplet must have a zero mode.
\item \textcolor{ForestGreen}{Green} corresponds to the two MSSM Higgses which are flavour triplets and must have zero modes.
\item \textcolor{violet}{Violet} correspond to right handed neutrinos and may or may not have a zero mode.
\item \textcolor{magenta}{Magenta} correspond to adjoint scalars that could obtain a VEV through the Wilson line. They may or may not have zero modes.
\item \textcolor{Dandelion}{Yellow} corresponds to a scalar that could get a VEV that would generate Majorana masses for RHNs. It may or may not have a zero mode.
\item \textcolor{red}{Red} corresponds to mirror families of the SM fermions. In general the number of massless blue modes minus the number of massless red modes must equal one.
\item \textcolor{orange}{Orange} correspond to Higgses that would force a GUT-scale $\mu$ term.  Without them there is no $\mu$ term and it must be generated dynamically. 
\item \textcolor{brown}{Brown} correspond to multiplets that could obtain a VEV through a Wilson line and break the remaining symmetry. They may or may not have zero modes.
\item {\bf Black} correspond to fields that must not exist at low energies. They are 4th and 5th families of fermions, leptoquarks and the triplets usually  the Higgses in GUTs. Either they do not have zero modes or the zero mode is a vector-like pair. 
\end{itemize}

A systematic scan of all the integers that define the orbifold was performed to find one that fulfills all the previous requirements. 
The whole parameter space for the orbifold was scanned for $N,M\leq 7$ (larger numbers did not yield a different result). While there are many setups for the integers, they are all physically equivalent. After finding a candidate solution we apply the extra constraint that there has to be a compatible lattice with the orbifolding and the remaining SUSY, as discussed in \cite{Fischer:2012qj}. We find that this leads to a unique choice of the orbifold $T^6/(\mathbb{Z}_6\times \mathbb{Z}_2)$.
In these cases solutions were found which 
satisfy the criteria that the field content of zero modes has three families of fermions.
However such solutions always have a larger 
Pati-Salam symmetry, as discussed in Appendix~\ref{app:orb}.
This motivates a dedicated analysis of an Exceptional Pati-Salam orbifold model in the next section.

\section{Exceptional Pati-Salam Model}
\label{sec:psorb}

\subsection{The orbifold $T^6/(\mathbb{Z}_6\times \mathbb{Z}_2)$}
In this section, a particular orbifold $T^6/(\mathbb{Z}_6\times \mathbb{Z}_2)$ is considered
where $\mathbb{Z}_6\in U(1)_{X'}$ as in subsection~\ref{SO(10)} and $\mathbb{Z}_2\in U(1)_{Y}$.
\footnote{For a non-SUSY Pati-Salam model see Appendix~\ref{nonsusy}.}

Under 
$E_8\to SU(4)_{PS}\times SU(2)_L\times SU(2)_R\times U(1)_{X'}\times SU(3)_F$
the decomposition of the adjoint representation is (using 
a similar colour coding as before):
\begin{equation}\begin{split}
(\textbf{248}) &\to \textcolor{magenta}{(\textbf{15},\textbf{1},\textbf{1},0,\textbf{1})}+  \textcolor{magenta}{(\textbf{1},\textbf{3},\textbf{1},0,\textbf{1})}+\textcolor{magenta}{(\textbf{1},\textbf{1},\textbf{3},0,\textbf{1})}+\textcolor{magenta}{(\textbf{1},\textbf{1},\textbf{1},0,\textbf{1})}+\textcolor{magenta}{(\textbf{1},\textbf{1},\textbf{1},0,\textbf{8})}\\
&\quad +(\textbf{6},\textbf{2},\textbf{2},0,\textbf{1})+ (\textbf{4},\textbf{2},\textbf{1},-3,\textbf{1})+ (\bar{\textbf{4}},\textbf{1},\textbf{2},-3,\textbf{1})+(\bar{\textbf{4}},\textbf{2},\textbf{1},3,\textbf{1})+(\textbf{4},\textbf{1},\textbf{2},3,\textbf{1}) \\
&\quad+ \textcolor{blue}{(\textbf{4},\textbf{2},\textbf{1},1,\textbf{3})}+ \textcolor{blue}{(\bar{\textbf{4}},\textbf{1},\textbf{2},1,\textbf{3})}+\textcolor{ForestGreen}{(\textbf{1},\textbf{2},\textbf{2},-2,\textbf{3})}+(\textbf{6},\textbf{1},\textbf{1},-2,\textbf{3})+\textcolor{brown}{(\textbf{1},\textbf{1},\textbf{1},4,\textbf{3})}\\
&\quad+ \textcolor{red}{(\bar{\textbf{4}},\textbf{2},\textbf{1},-1,\bar{\textbf{3}})}+ \textcolor{red}{(\textbf{4},\textbf{1},\textbf{2},-1,\bar{\textbf{3}})} +\textcolor{orange}{(\textbf{1},\textbf{2},\textbf{2},2,\bar{\textbf{3}})}+(\textbf{6},\textbf{1},\textbf{1},2,\bar{\textbf{3}})+\textcolor{brown}{(\textbf{1},\textbf{1},\textbf{1},-4,\bar{\textbf{3}})}.
\end{split}
\label{PS}
\end{equation}
The same color coding as in eq. \ref{eq:e8sm} has been used, with the difference that some fields that were black are now part of the adjoints in magenta. Also the right handed neutrinos that were in violet 
${\nu}^c\sim \textcolor{violet}{(\textbf{1},\textbf{1},0,-5,1,\textbf{3})}$
are now part of the two PS fermion multiplets in blue (and their conjugate representations in yellow $\bar{\nu}^c\sim \textcolor{YellowOrange}{(\textbf{1},\textbf{1},0,5,-1,\bar{\textbf{3}})}$
now being part of the red ones).

This decomposition corresponds to that of eq.~\eqref{eq:so10rep} with the further breaking 
\begin{equation}
SO(10)\rightarrow SU(4)_{PS}\times SU(2)_L\times SU(2)_R,
\label{so10breaking}
\end{equation}
\begin{equation}\begin{split}
\textbf{10}\rightarrow (\textbf{6},\textbf{1},\textbf{1}) + (\textbf{1},\textbf{2},\textbf{2}),\ \ 
\textbf{16}\rightarrow (\textbf{4},\textbf{2},\textbf{1}) + (\bar{\textbf{4}},\textbf{1},\textbf{2}),\\
\textbf{45}\rightarrow (\textbf{15},\textbf{1},\textbf{1}) + (\textbf{1},\textbf{3},\textbf{1})
+(\textbf{1},\textbf{1},\textbf{3})+ (\textbf{6},\textbf{2},\textbf{2}).
\end{split}\end{equation}
The SM group would be obtained from PS by the subsequent breakings
\begin{equation}
SU(4)_{PS}\rightarrow SU(3)_C\times U(1)_{B-L},
\end{equation}
\begin{equation}\begin{split}
\textbf{4}\rightarrow (\textbf{3},1/3) + (\textbf{1},-1),\ \  
\textbf{6}\rightarrow (\textbf{3},-2/3) + (\bar{\textbf{3}},2/3),\\
\textbf{15}\rightarrow  (\textbf{8},0)+  (\textbf{1},0) + (\textbf{3},4/3)
+ (\bar{\textbf{3}},-4/3)
\end{split}\end{equation}
and
\begin{equation}
SU(2)_R \rightarrow U(1)_{T^R_3},
\end{equation}
\begin{equation}
\textbf{2}\rightarrow (1/2) + (-1/2),\ \  \textbf{3}\rightarrow (1) + (0) +(-1),
\end{equation}
which together would yield the QCD and hypercharge gauge groups~\cite{King:2017cwv}
\begin{equation}
SU(4)_{PS}\times SU(2)_R \rightarrow SU(3)_C\times U(1)_{Y} \times U(1)_{X},
\label{PSbreaking}
\end{equation}
with the decomposition
\begin{equation}\begin{split}
(\textbf{4}, \textbf{1})\rightarrow (\textbf{3},1,-1) + (\textbf{1},-3,3),\\  
(\textbf{6}, \textbf{1})\rightarrow (\textbf{3},-2,2) + (\bar{\textbf{3}},2,-2),\\  
(\textbf{15}, \textbf{1})\rightarrow (\textbf{8},0,0) + (\textbf{1},0,0) + (\textbf{3},4,-4) +(\bar{\textbf{3}},-4,4),\\  
(\textbf{1}, \textbf{2})\rightarrow (\textbf{1},3,2) + (\textbf{1},-3,-2),\\  
(\textbf{1}, \textbf{3})\rightarrow (\textbf{1},6,4) + (\textbf{1},0,0) + (\textbf{1},-6,-4),\\  
(\textbf{4}, \textbf{2})\rightarrow (\textbf{3},4,1) + (\textbf{1},0,5) + (\textbf{3},-2,-3) + (\textbf{1},-6,1),\\
(\textbf{6}, \textbf{2})\rightarrow (\textbf{3},1,4) + (\bar{\textbf{3}},5,0) + (\textbf{3},-5,0) + (\bar{\textbf{3}},-1,-4),
\end{split}\end{equation}
where in our normalisation $Y=6T^R_3+3(B-L)$ and $X=4T^R_3-3(B-L)$.

The $T^6/(\mathbb{Z}_6\times \mathbb{Z}_2)$ orbifold which achieves the breaking in eq.~\ref{PS} is defined as:
\footnote{This is equivalent to one of the cases tabulated in~\cite{Fischer:2012qj}.}
\begin{equation}
\mathbb{Z}_6:\ \phi \to e^{2i\pi q_{X'}/6}\phi,\ \ \  \mathbb{Z}_2:\ \phi \to e^{2i\pi q_{Y}/2}\phi,
\end{equation}
\begin{equation}\begin{split}
(x,z_1,z_2,z_3) &\sim (x,\alpha^2 z_1,\alpha^5 z_2, \alpha^5 z_3),\\
(x,z_1,z_2,z_3) &\sim (x, -z_1,- z_2, (-1)^2 z_3),
\label{eq:orbrot}
\end{split}\end{equation}
where $\alpha=e^{2i\pi/6}$ and $-1=e^{2i\pi/2}$. The $(-1)^2$ implies a full rotation on the $z_3$ to comply with eq. \ref{eq:ni}.

The $\mathbb{Z}_6$ orbifolding is based on 
$U(1)_{X'}$ which commutes with $SO(10)$ and $SU(3)_F$. This orbifold operation breaks $E_8\to SO(10)\times U(1)_{X'}\times SU(3)_F$, as can be seen in table~\ref{tab:e8} and subsection~\ref{SO(10)}.
The $\mathbb{Z}_2$ orbifolding is based on $U(1)_{Y}$, the hypercharge. 
This orbifolding breaks $E_8\to E_7\times SU(2)$ as seen in table \ref{tab:e8}. This $E_7$ does not contain the previous $SO(10)$ and its intersection is the Pati-Salam group. 
The breaking of $SO(10)$ in eq.~\ref{so10breaking} occurs since all the 
gauge bosons in PS have even hypercharge parity $\mathbb{Z}_2$, unlike the $\textbf{45}$ of $SO(10)$
which contains $(\textbf{6},\textbf{2},\textbf{2})$ which has odd hypercharge parity $\mathbb{Z}_2$.
The orbifolds charges 
of all the resulting $\mathcal{N}=1$ superfields can be seen in table~\ref{tab:pps}.

The massless zero modes in this Pati-Salam setup are only the $\mathcal{N}=1$ superfields
with both charges equal to unity (the singlets $1,1$) from table \ref{tab:pps}:
\begin{equation}\begin{split}
V_\mu &: \textcolor{magenta}{(\textbf{15},\textbf{1},\textbf{1},0,\textbf{1})}+  \textcolor{magenta}{(\textbf{1},\textbf{3},\textbf{1},0,\textbf{1})}+\textcolor{magenta}{(\textbf{1},\textbf{1},\textbf{3},0,\textbf{1})}+\textcolor{magenta}{(\textbf{1},\textbf{1},\textbf{1},0,\textbf{1})}+\textcolor{magenta}{(\textbf{1},\textbf{1},\textbf{1},0,\textbf{8})},\\
\phi_1&:\textcolor{ForestGreen}{(\textbf{1},\textbf{2},\textbf{2},-2,\textbf{3})},\\
\phi_2&: \textcolor{blue}{(\textbf{4},\textbf{2},\textbf{1},1,\textbf{3})}, \\
\phi_3&:\textcolor{blue}{(\bar{\textbf{4}},\textbf{1},\textbf{2},1,\textbf{3})}.\\
\label{eq:zmf}
\end{split}\end{equation}

\begin{table}
	\centering
	\scriptsize
	\renewcommand{\arraystretch}{1.1}
	\begin{tabular}[t]{l|llll}
		\hline
		 & $V$ & $\phi_1$ & $\phi_2$ & $\phi_3$\\ 
		\hline
	$\mathcal{V}_{\textcolor{magenta}{(\textbf{15},\textbf{1},\textbf{1},0,\textbf{1})}} $ & $1,1$ & $\alpha^2, -1$ & $\alpha^5, -1$& $\alpha^5, 1$\\
		$\mathcal{V}_{\textcolor{magenta}{(\textbf{1},\textbf{3},\textbf{1},0,\textbf{1})}} $ & $1,1$ & $\alpha^2, -1$ & $\alpha^5, -1$& $\alpha^5, 1$\\
		$\mathcal{V}_{\textcolor{magenta}{(\textbf{1},\textbf{1},\textbf{3},0,\textbf{1})}} $ & $1,1$ & $\alpha^2, -1$ & $\alpha^5, -1$& $\alpha^5, 1$\\
		$\mathcal{V}_{\textcolor{magenta}{(\textbf{1},\textbf{1},\textbf{1},0,\textbf{1})}} $ & $1,1$ & $\alpha^2, -1$ & $\alpha^5, -1$& $\alpha^5, 1$\\
		$\mathcal{V}_{\textcolor{magenta}{(\textbf{1},\textbf{1},\textbf{1},0,\textbf{8})}} $ & $1,1$ & $\alpha^2, -1$ & $\alpha^5, -1$& $\alpha^5, 1$\\
		$\mathcal{V}_{(\textbf{6},\textbf{2},\textbf{2},0,\textbf{1})} $& $1,-1$ & $\alpha^2, 1$ & $\alpha^5, 1$& $\alpha^5, -1$\\
		$\mathcal{V}_{(\textbf{4},\textbf{2},\textbf{1},-3,\textbf{1})} $  & $\alpha^3,-1$ & $\alpha^5, 1$ & $\alpha^2, 1$& $\alpha^2, -1$\\
		$\mathcal{V}_{(\bar{\textbf{4}},\textbf{1},\textbf{2},-3,\textbf{1})} $ & $\alpha^3,1$ & $\alpha^5, -1$ & $\alpha^2, -1$& $\alpha^2, 1$\\
		$\mathcal{V}_{(\bar{\textbf{4}},\textbf{2},\textbf{1},3,\textbf{1})} $  & $\alpha^3,-1$ & $\alpha^5, 1$ & $\alpha^2, 1$& $\alpha^2, -1$\\		
		$\mathcal{V}_{(\textbf{4},\textbf{1},\textbf{2},3,\textbf{1})} $ & $\alpha^3,1$ & $\alpha^5, -1$ & $\alpha^2, -1$& $\alpha^2, 1$ \\	
		\hline
	\end{tabular}
	\hspace*{0.3cm}
	\begin{tabular}[t]{l|llll}
		\hline
		 & $V$ & $\phi_1$ & $\phi_2$ & $\phi_3$\\ 
		\hline
	$\mathcal{V}_{\textcolor{blue}{(\textbf{4},\textbf{2},\textbf{1},1,\textbf{3})}} $ & $\alpha,
	-1$ & $\alpha^3, 1$ & $1, 1$& $1, -1$\\
		$\mathcal{V}_{\textcolor{blue}{(\bar{\textbf{4}},\textbf{1},\textbf{2},1,\textbf{3})}} $ & $\alpha,1$ & $\alpha^3, -1$ & $1, -1$& $1, 1$\\
		$\mathcal{V}_{\textcolor{ForestGreen}{(\textbf{1},\textbf{2},\textbf{2},-2,\textbf{3})}} $  & $\alpha^4,-1$ & $1, 1$ & $\alpha^3, 1$& $\alpha^3, -1$\\
		$\mathcal{V}_{(\textbf{6},\textbf{1},\textbf{1},-2,\textbf{3})} $ & $\alpha^4,1$ & $1, -1$ & $\alpha^3,- 1$& $\alpha^3, 1$\\
		$\mathcal{V}_{\textcolor{brown}{(\textbf{1},\textbf{1},\textbf{1},4,\textbf{3})}} $ & $\alpha^4,1$ & $1, -1$ & $\alpha^3, -1$& $\alpha^3, 1$\\
		$\mathcal{V}_{\textcolor{red}{(\bar{\textbf{4}},\textbf{2},\textbf{1},-1,\bar{\textbf{3}})}} $& $\alpha^5,-1$ & $\alpha,1$ & $\alpha^4, 1$& $\alpha^4, -1$\\
		$\mathcal{V}_{\textcolor{red}{(\textbf{4},\textbf{1},\textbf{2},-1,\bar{\textbf{3}})} } $  & $\alpha^5,1$ & $\alpha, -1$ & $\alpha^4, -1$& $\alpha^4, 1$\\
		$\mathcal{V}_{\textcolor{orange}{(\textbf{1},\textbf{2},\textbf{2},2,\bar{\textbf{3}})}} $ & $\alpha^2,-1$ & $\alpha^4, 1$ & $\alpha, 1$& $\alpha, -1$\\
		$\mathcal{V}_{(\textbf{6},\textbf{1},\textbf{1},2,\bar{\textbf{3}})} $  & $\alpha^2,1$ & $\alpha^4, -1$ & $\alpha,- 1$& $\alpha, 1$\\		
		$\mathcal{V}_{\textcolor{brown}{(\textbf{1},\textbf{1},\textbf{1},-4,\bar{\textbf{3}})}} $  & $\alpha^2,1$ & $\alpha^4, -1$ & $\alpha, -1$& $\alpha, 1$\\	
		\hline
	\end{tabular}
	\caption{$\mathbb{Z}_6\times \mathbb{Z}_2$ orbifold charges of each  $\mathcal{N}=1$ superfield. Only the superfields with both charges equal to unity 
	(the singlets $1,1$) have zero modes.} 
	\label{tab:pps}
\end{table}

The massless superfields can be named as
\begin{equation} \label{eq:phis}
\begin{split}
V_\mu &: \textcolor{magenta}{G_\mu}+\textcolor{magenta}{W^L_\mu}+\textcolor{magenta}{W^R_\mu}+\textcolor{magenta}{Z'_\mu}+\textcolor{magenta}{F_\mu},\\
\phi_1&: \textcolor{ForestGreen}{H},\\
\phi_2&: \textcolor{blue}{F},\\
\phi_3&:\textcolor{blue}{F^c}.
\end{split}\end{equation}
The chiral superfields decompose into the SM fields as 
\begin{equation}\begin{split}
\textcolor{ForestGreen}{H}&\to \textcolor{ForestGreen}{h_u}+\textcolor{ForestGreen}{h_d},\\
\textcolor{blue}{F}&\to \textcolor{blue}{Q}+\textcolor{blue}{L},\\
\textcolor{blue}{F^c}&\to \textcolor{blue}{u^c}+\textcolor{blue}{d^c}+\textcolor{Violet}{\nu^c}+\textcolor{blue}{e^c}.
\end{split}\end{equation}

The SM gauge group is subsequently achieved by appealing to Wilson line breaking of the PS and other gauge group factors in addition to the orbifold breaking.
The resulting model is dubbed the Exceptional Pati-Salam (EPS) model.

The only renormalizable term in the superpotential from eq.~\eqref{eq:n41lag} becomes
\begin{equation}
\mathcal{W}\sim \phi_1\phi_2\phi_3=HF F^c,
\end{equation}
which are the Yukawa couplings.

In summary, $E_8$ has been broken down to Pati-Salam with an $SU(3)_F$ flavour group. 
Starting from the bulk 
state $\mathcal{V}_{\textbf{248}}\sim  \textbf{248}$, the low energy
field content
is that of the SM apart from the SM singlet right handed neutrinos, and extra Higgs doublets forming flavour triplets. There are no other fields with zero modes, i.e. no mirror fermions, exotics and effective doublet-triplet splitting .
In order to achieve three (and only three) chiral superfields with the correct SM field content required at least 10d, 
where this conclusion is unrelated to String Theory.

This is the unique orbifold that eliminates mirror fermions, solves the DT splitting and preserves simple SUSY \cite{Fischer:2012qj,Parr:2020oar}.

\subsection{Anomaly cancellation}
\label{sec:anom}
The zero modes listed in eq. \ref{eq:zmf} are not anomaly free.
The fermion components have chiral gauge anomalies associated with one or more of the gauge groups
$U(1)_{X'}\times SU(3)_F$.
To cancel such anomalies we shall add extra states which do not spoil the good high energy behaviour of the theory,
and which are also consistent with our approach that the SM fields arise solely from the bulk 
state $\mathcal{V}_{\textbf{248}}\sim  \textbf{248}$. 

The choice of the extra field content to cancel the anomalies is not unique. We present the simplest possible being consistent at high energies while not affecting the low energy theory. 

In order to cancel anomalies which involve the gauge group $U(1)_{X'}$, we shall require 
two extra massless mode chiral multiplets, namely 
$(\textbf{6},\textbf{1},\textbf{1},-2,\textbf{3})$,  and $\textcolor{brown}{(\textbf{1},\textbf{1},\textbf{1},4,\textbf{3})}$.
Together with the massless modes in eq. \ref{eq:zmf}, it is readily checked that these form an anomaly free set of states
(apart from anomalies involving $SU(3)_F$). The reason is that the $\textbf{27}$ of $E_6$ is anomaly free and hence the subgroup
$SU(4)_{PS}\times SU(2)_L\times SU(2)_R\times U(1)_{X'}$ containing all the decomposed states of the $\textbf{27}$
will also be anomaly free.
The extra states may arise, for example, from an extra 6d superfield ${\phi}_{\textbf{248}}^{(z_1)}$ located in the $z_1$ torus, 
which is assigned orbifold charges of $(1,-1)$, such that the zero modes 
are those with charge $(1,-1)$ from the $\phi_1$ column in table \ref{tab:pps}.
This yields the extra chiral multiplets in
$(\textbf{6},\textbf{1},\textbf{1},-2,\textbf{3})+\textcolor{brown}{(\textbf{1},\textbf{1},\textbf{1},4,\textbf{3})}$,
which complete the decomposed $\textbf{27}$ representation.

The above two extra massless mode chiral multiplets $(\textbf{6},\textbf{1},\textbf{1},-2,\textbf{3})$,  and $\textcolor{brown}{(\textbf{1},\textbf{1},\textbf{1},4,\textbf{3})}$ may alternatively arise from a set of 10d chiral superfields.
Assuming that $\mathcal{N}=1$ SUSY is preserved in the 4d theory, 
a $10$d chiral superfield $\mathcal{\phi}_{\textbf{248}}$ 
decomposes into four $4$d chiral superfield multiplets $\phi_{0,1,2,3}$,
\begin{equation}
\mathcal{\phi}_{\textbf{248}} \rightarrow \{\phi_{0,1,2,3}\}_{\textbf{248}},
\label{scalar}
\end{equation}
where the $\mathcal{N}=1$ SUSY chiral scalar superfields $\phi_{0,1,2,3}$ are the analogue of the extra dimensional components of the 
10d gauge field $\mathcal{V}$ in eq.\eqref{vector}. The $\mathcal{\phi}_{\textbf{248}}$ can be assigned arbitrary orbifold charges, such that the massless modes can be read off from table \ref{tab:pps}, with $\phi_0$ replacing $V$, and the surviving massless modes corresponding to the inverse of the orbifold charge assigned to $\mathcal{\phi}_{\textbf{248}}$. Then it can be readily shown that a collection of five $10$d chiral superfields $\mathcal{\phi}_{\textbf{248}}$ with the surviving massless modes given by the orbifold charges $(1,\pm1),(\alpha^4,\pm 1),(\alpha^2,-1)$, respectively, will yield 
the desired two extra massless mode chiral multiplets $(\textbf{6},\textbf{1},\textbf{1},-2,\textbf{3})$,  and $\textcolor{brown}{(\textbf{1},\textbf{1},\textbf{1},4,\textbf{3})}$, plus additional states which are either real or vector-like under the gauge group, leading to
anomaly cancellation.

At this stage, we have $27$ $SU(3)_F$ net triplets (plus an equal number of triplet-antitriplet pairs). Without further additions, 
this theory would still suffer from an $SU(3)_F-SU(3)_F-SU(3)_F$ gauge anomaly, which must be cancelled. In 
string theory constructions, this gauge anomaly is typically cancelled by $27$ anti-triplets of $SU(3)_F$ that are localised at some fixed points of the orbifold. So it is natural to assume that these anomalies are cancelled by $27$ anti-triplets of $SU(3)_F$, which are $E_6$ singlets,
located at a 4d brane, for example at $z_1=z_2=z_3=0$, which respects $E_6\times SU(3)_{F}$ but not the full $E_8$
(see for example~\cite{Katsuki:1989kd}).
Such states are $E_6$ singlets, and are expected  
to gain mass at the $SU(3)_F$ breaking scale.
The extra states introduced in this section are consistent with our approach that all SM matter arises from the bulk 
state $\mathcal{V}_{\textbf{248}}\sim  \textbf{248}$.
Moreover, now that the theory is anomaly free we may consistently consider further Wilson line breaking of the gauge symmetry.

\subsection{Wilson line breaking of $E_8$ }

The orbifolding has broken $E_8\to SU(4)_{PS}\times SU(2)_L\times SU(2)_R\times U(1)_{X'}\times SU(3)_F$, which is still rank $8$. Orbifolding by itself cannot reduce the rank and Wilson lines are needed to further break the 
PS symmetry as in eq.~\eqref{PSbreaking} and also to break $U(1)_{X}\times U(1)_{X'}\times SU(3)_F$.

We first recall that the right handed neutrinos
${\nu}^c\sim \textcolor{violet}{(\textbf{1},\textbf{1},0,-5,1,\textbf{3})}$ in eq.~\eqref{eq:e8sm} 
are part of the PS fermion multiplet
$F^c\sim \textcolor{blue}{(\bar{\textbf{4}},\textbf{1},\textbf{2},1,\textbf{3})}$. The scalar components $\tilde{\nu}^c$, being singlets under the SM symmetry, therefore makes them candidates to obtain VEVs through a Wilson line, where such VEVs 
are naturally at the compactification scale.

The VEV of the scalar component of the RH neutrino 
$\tilde{\nu}^c \sim \textcolor{violet}{(\textbf{1},\textbf{1},0,-5,1,\textbf{3})}$
through a Wilson line in $\phi_2$, will be determined via 
some effective potential \cite{Haba:2002py,Hosotani:2004ka}.
Since $\mathcal{N}=1$ SUSY is preserved after compactification, presumably the VEV in the 
scalar component of the complex conjugated representation in eq.~\eqref{eq:e8sm}
$\tilde{\bar{\nu}}^c \sim \textcolor{YellowOrange}{(\textbf{1},\textbf{1},0,5,-1,\bar{\textbf{3}})}$,
would also be induced due to D flatness, even though it is a massive KK mode. 

The two VEVs $\langle \tilde{\nu}^c \rangle $ and $\langle  \tilde{\bar{\nu}}^c \rangle $ break  
\begin{equation}
SU(4)_{PS}\times SU(2)_L\times SU(2)_R\times U(1)_{X'}\times SU(3)_F
\rightarrow SU(3)_{C}\times SU(2)_L\times U(1)_Y\times U(1)_N,
\end{equation}
where $U(1)_N$ is the $E_6$ subgroup under which the RH neutrinos are neutral~\cite{King:2005jy}, where in our convention, the 
corresponding generator is given by 
\begin{equation}
 N=X+5{X'}.
 \end{equation}
It is interesting to note that at this stage the matter content is identical to that of the 
E$_6$SSM~\cite{King:2005jy,Howl:2007zi,King:2020ldn}, comprising the superfields in 
eq.~\eqref{eq:zmf} plus the additional superfields in $(\textbf{6},\textbf{1},\textbf{1},-2,\textbf{3})$,  and 
$\textcolor{brown}{(\textbf{1},\textbf{1},\textbf{1},4,\textbf{3})}$, required for anomaly cancellation.
To be precise, this is the matter content of the minimal E$_6$SSM~\cite{Howl:2007zi}, 
comprising three 27 component families, which here is achieved via an intermediate PS gauge group.
In this case, all the above matter states may appear at the few TeV scale, where the $U(1)_N$ is broken via a VEV in the scalar component of the singlets arising from the $\textcolor{brown}{(\textbf{1},\textbf{1},\textbf{1},4,\textbf{3})}$, as in the E$_6$SSM.

However it is also possible that the $U(1)_N$ may broken at high energies by VEVs in the scalar components of 
$\textcolor{brown}{(\textbf{1},\textbf{1},\textbf{1},4,\textbf{3})}$. In this case 
we would also expect a VEV from $\textcolor{brown}{(\textbf{1},\textbf{1},\textbf{1},-4,\bar{\textbf{3}})}$, in order to preserve D flatness, even though it is a massive KK mode.
In this case,
with only the SM gauge group surviving, there would no longer be any symmetry which can protect states which are singlets or
vector-like under it from gaining a mass, including the extra states which were required to cancel
gauge anomalies, which will not be present in the low energy spectrum.
At low energies this suggests the following superfields of the SM gauge group $SU(3)_C\times SU(2)_L\times U(1)_Y$:
 \begin{equation}\begin{split}
V_\mu &:  \textcolor{magenta}{(\textbf{8},\textbf{1},0)}+\textcolor{magenta}{(\textbf{1},\textbf{3},0)}+\textcolor{magenta}{(\textbf{1},\textbf{1},0)},\\
\phi_1&:3\times\textcolor{ForestGreen}{(\textbf{1},\textbf{2},3)}+3\times\textcolor{ForestGreen}{(\textbf{1},\textbf{2},-3)},\\
\phi_2&: 3\times\textcolor{blue}{(\textbf{3},\textbf{2},1)}+ 3\times\textcolor{blue}{(\textbf{1},\textbf{2},-3)}, \\
\phi_3&:3\times\textcolor{blue}{(\textbf{1},\textbf{1},6)}+3\times\textcolor{blue}{(\bar{\textbf{3}},\textbf{1},-4)}+3\times\textcolor{blue}{(\bar{\textbf{3}},\textbf{1},2)}+3\times \textcolor{violet}{(\textbf{1},\textbf{1},0)},
\label{eq:zmff}
\end{split}\end{equation}
which can be named as
   \begin{equation}\begin{split}
V_\mu &:  \textcolor{magenta}{G_\mu}+\textcolor{magenta}{W_\mu}+\textcolor{magenta}{B_\mu},\\
\phi_1&:\textcolor{ForestGreen}{h_{ui}}+\textcolor{ForestGreen}{h_{di}},\\
\phi_2&: \textcolor{blue}{Q_i}+\textcolor{blue}{L_i}, \\
\phi_3&:\textcolor{blue}{e^c_i}+\textcolor{blue}{u^c_i}+\textcolor{blue}{d^c_i}+ \textcolor{violet}{\nu^c_i},
\end{split}\end{equation}
  with $i=1,2,3$. This is the MSSM field content, plus 3 right handed neutrinos plus 2 extra pairs of Higgs doublets.
  Henceforth we shall assume this low energy spectrum.

\subsection{Effects of the right-handed sneutrino VEV}
The effective Yukawa terms become (up to Clebsch-Gordan coefficients)
\begin{equation}
\mathcal{W}_Y\sim F H F^c+ F H F^c(\bar{\nu}^c\nu^c)+(F\bar{\nu}^c)(\nu^c H F^c)+(F H \nu^c)(\bar{\nu}^cF^c)+...
\end{equation}
where the higher order terms are mediated by KK modes with corresponding mass scales not shown. 
Assuming that both the RH sneutrinos and the 
Higgses get a VEV $\braket{h^{u,d}_i}=v^{u,d}_i$,
these terms generate the mass matrices 
\begin{equation}
\begin{split}
M_{u,\nu,e,d}&\sim v^{u,d}_1\left[\left(\begin{array}{ccc} 0 & 0 & 0\\
0&0& -1 -\braket{\tilde{\bar{\nu}}^c_{i}\tilde{\nu}^c_{i}} \\
0  & 1 +\braket{\tilde{\bar{\nu}}^c_{i}\tilde{\nu}^c_{i}} &  0
\end{array}\right)+\left(\begin{array}{ccc} 0 & \braket{\tilde{\bar{\nu}}^c_{1}\tilde{\nu}^c_{3}} & -\braket{\tilde{\bar{\nu}}^c_{1}\tilde{\nu}^c_{2}}\\
\braket{\tilde{\bar{\nu}}^c_{1}\tilde{\nu}^c_{3}} & \braket{\tilde{\bar{\nu}}^c_{2}\tilde{\nu}^c_{3}} &0  \\
-\braket{\tilde{\bar{\nu}}^c_{1}\tilde{\nu}^c_{2}}  & 0 &  -\braket{\tilde{\bar{\nu}}^c_{3}\tilde{\nu}^c_{2}}
\end{array}\right)\right]\\
 &\quad+ v^{u,d}_2\left[\left(\begin{array}{ccc} 0 & 0 & 1+ \braket{\tilde{\bar{\nu}}^c_{i}\tilde{\nu}^c_{i}} \\
0 & 0 & 0\\
-1 -\braket{\tilde{\bar{\nu}}^c_{i}\tilde{\nu}^c_{i}}  & 0 &  0
\end{array}\right)+\left(\begin{array}{ccc} \braket{\tilde{\bar{\nu}}^c_{1}\tilde{\nu}^c_{3}}  & -\braket{\tilde{\bar{\nu}}^c_{2}\tilde{\nu}^c_{3}} &0\\
-\braket{\tilde{\bar{\nu}}^c_{2}\tilde{\nu}^c_{3}}& 0 & \braket{\tilde{\bar{\nu}}^c_{2}\tilde{\nu}^c_{1}} \\
0 &\braket{\tilde{\bar{\nu}}^c_{2}\tilde{\nu}^c_{1}}&  \braket{\tilde{\bar{\nu}}^c_{3}\tilde{\nu}^c_{1}}
\end{array}\right)\right]\\
&\quad+ v^{u,d}_3\left[\left(\begin{array}{ccc}0  & -1  -\braket{\tilde{\bar{\nu}}^c_{i}\tilde{\nu}^c_{i}}& 0 \\
1 +\braket{\tilde{\bar{\nu}}^c_{i}\tilde{\nu}^c_{i}}&0& 0\\
0 & 0 &0
\end{array}\right)+\left(\begin{array}{ccc}\braket{\tilde{\bar{\nu}}^c_{1}\tilde{\nu}^c_{2}} &0 & \braket{\tilde{\bar{\nu}}^c_{3}\tilde{\nu}^c_{2}}\\
0&\braket{\tilde{\bar{\nu}}^c_{2}\tilde{\nu}^c_{1}}& -\braket{\tilde{\bar{\nu}}^c_{3}\tilde{\nu}^c_{1}}\\
\braket{\tilde{\bar{\nu}}^c_{3}\tilde{\nu}^c_{2}}  & -\braket{\tilde{\bar{\nu}}^c_{3}\tilde{\nu}^c_{1}}&0
\end{array}\right)\right].
\end{split}
\end{equation}
Assuming that the VEVs are of order the KK mass scale leads to 
a democratic structure. Even though there are no available free Yukawa couplings from the original theory (there is only one gauge coupling and fixed Clebsch-Gordan coefficients), they can be effectively generated by the VEVs $v^{u,d},\braket{\tilde{\nu}^c},\braket{\tilde{\bar{\nu}}^c}$ which introduce eleven complex free parameters to fit the flavour data. To reproduce the hierarchical pattern of SM fermion masses and mixing angles will require some tuning due to the
democratic structure.

A VEV in the right handed sneutrino generates the 
effective bilinear R-parity violating term (since the chiral superfields $\phi_{2,3}$ have odd R-parity) 
\begin{equation} \label{eq:RPV-term}
L h_u\braket{\tilde{\nu}^c}.
\end{equation}
Assuming an even larger Higgs mass superpotential term $\mu h_u h_d$, where $\mu > \braket{\tilde{\nu}^c}$
(the various scales are discussed in the next subsection)
the effective bilinear R-parity violating term in Eq.\ref{eq:RPV-term} can be rotated into the Yukawa matrices 
by a unitary transformation~\cite{Dreiner:1997uz}.
\begin{equation} \label{eq:RPV-rotation}
(L,h_d)\to V^a_{ij} (L,h_d),
\end{equation}
where $a=1,2$ for $L$ and $h_d$ respectively, and $i,j=1,2,3$. The rotation would only leave R-parity violating terms in the Yuakawa sector. It is very important to note that this rotation does not affect the up quark mass matrix nor the charged lepton mass matrix, since it is unitary and it involves both $L$ and $h_d$.

This rotation $V$ however, changes the down quark mass matrix by changing
\begin{equation}
v^d_j\to v^d_i V^2_{ij},
\end{equation}
therefore generating a difference between the down quark mass matrix and unchanged the charged lepton mass matrix. Furthermore
it also changes the neutrino mass matrix 
\begin{equation}
M_\nu \to V^{1} M_\nu ,
\end{equation}
and thus the VEV $\braket{\tilde{\nu}^c}$ breaks the universal mass matrix structure, giving a pathway for a realistic set of SM fermion mass matrices (including neutrinos). 

 This VEVs would generate a right handed neutrino Majorana mass term
  \begin{equation}
  \nu^c\nu^c\braket{\tilde{\bar{\nu}}^c}\braket{\tilde{\bar{\nu}}^c},
  \end{equation}
  and the Majorana mass matrix
  \begin{equation}
  M_{NN}\sim \left(\begin{array}{ccc}
  2\braket{\tilde{\bar{\nu}}^c_1\tilde{\bar{\nu}}^c_1} &  \braket{\tilde{\bar{\nu}}^c_1\tilde{\bar{\nu}}^c_2}+ \braket{\tilde{\bar{\nu}}^c_2\tilde{\bar{\nu}}^c_1} &  \braket{\tilde{\bar{\nu}}^c_1\tilde{\bar{\nu}}^c_3}+ \braket{\tilde{\bar{\nu}}^c_3\tilde{\bar{\nu}}^c_1}\\
   \braket{\tilde{\bar{\nu}}^c_2\tilde{\bar{\nu}}^c_1}+ \braket{\tilde{\bar{\nu}}^c_1\tilde{\bar{\nu}}^c_2} &  2\braket{\tilde{\bar{\nu}}^c_2\tilde{\bar{\nu}}^c_2} &  \braket{\tilde{\bar{\nu}}^c_2\tilde{\bar{\nu}}^c_3}+ \braket{\tilde{\bar{\nu}}^c_3\tilde{\bar{\nu}}^c_2}\\
    \braket{\tilde{\bar{\nu}}^c_1\tilde{\bar{\nu}}^c_3}+ \braket{\tilde{\bar{\nu}}^c_3\tilde{\bar{\nu}}^c_1} &  \braket{\tilde{\bar{\nu}}^c_3\tilde{\bar{\nu}}^c_2}+ \braket{\tilde{\bar{\nu}}^c_2\tilde{\bar{\nu}}^c_3} &  2\braket{\tilde{\bar{\nu}}^c_3\tilde{\bar{\nu}}^c_3}
  \end{array}\right),
  \end{equation}
that can generate small left handed neutrino masses through the type-1 seesaw mechanism.

Finally one could be worried about the fact that the rotation generates the R-parity violating terms
\begin{equation}
(V^{1}v^d )Q L d^c.
\end{equation}
Fortunately the bounds associated with them are easily satisfied for a large SUSY breaking scale,
and also typically they are of the order of the down quark Yukawa couplings~\cite{Dreiner:1997uz}. It is important to note also that these terms by themselves do not induce proton decay.

In summary, the VEV $\braket{\tilde{\nu}^c}$ induced through a Wilson line can break the remaining symmetry into the SM, break the universality of masses and give Majorana masses to right handed neutrinos. This is at the cost of introducing R-parity violating terms which. however, should be naturally well below the experimental bounds. 

Lastly, since SUSY has been respected by the orbifold, the R-symmetry breaking due to $\braket{\tilde{\nu}^c}$ would induce SUSY breaking~\cite{Nelson:1993nf}. 

It seems that the existence of right-handed neutrinos, within the context of this fully unified setup, turns out to be crucial since it plays a fundamental role in the whole process. Usually, right-handed neutrinos are included to fill in representations and/or to help generate light masses for the observed neutrinos but have limited theoretical/phenomenological relevance beyond that. In this setup, they become essential, leading to experimental signatures, as discussed in the next subsection.

\subsection{Experimental Signatures}

As discussed above, this potentially troublesome R-parity violating term in Eq.~\eqref{eq:RPV-term} can be rotated
into the Yukawa sector by the unitary transformation in~\eqref{eq:RPV-rotation} so that it leaves R-parity violating terms only in the d-type quark and Higgs fields. However it was already noted that such a rotation presumes a larger Higgs $\mu$ term.
In this subsection, the various symmetry breaking scales and experimental signatures in the EPS model are discussed.

First note that when the right-handed sneutrino gets a VEV, the R-parity violating term in~\eqref{eq:RPV-term} leads to a Dirac mass term coupling the $L$ to the $h_u$. In the absence of other mass terms, these states will therefore be very heavy 
(close to the Pati-Salam breaking scale $\braket{\tilde{\nu}^c}$) and will thus decouple from the low energy theory. To make sure this does not happen in the model, it is necessary to consider a larger Higgs mass $\mu h_u h_d$, where $\mu > \braket{\tilde{\nu}^c}$ term. In order to achieve correct electroweak symmetry breaking this further implies a similarly large 
SUSY breaking scale $M_{SUSY}\sim O(\mu) $. 
Since the current bound on the Pati-Salam breaking scale $\braket{\tilde{\nu}^c}$ from the non-observation of $K_L\rightarrow \mu e$,
$n-\overline{n}$ oscillations and $B_{d,s}\rightarrow \mu e$
is around $10^6$ GeV~\cite{Bandyopadhyay:2015fka}, it must be assumed that $\mu$ is larger than that value (i.e. SUSY breaking must happen around the Pati-Salam breaking scale). Note that an effective $\mu$ term must be obtained radiatively. Putting all this together one is led to the suggestive pattern of scales in the EPS model,
\begin{equation}
M_{SUSY}\sim O (\mu)  \gtrsim \braket{\tilde{\nu}^c} \gtrsim 10^6 \ {\mathrm GeV}.
\end{equation}
While Wilson line breaking would suggest that $\braket{\tilde{\nu}^c}$ be very large, of order the compactification scale,
the imposed requirement that SUSY is preserved in the low energy theory, suggests that some compromise should be achieved with 
the PS breaking scale near its experimental limit $\braket{\tilde{\nu}^c} \sim 10^6 \ {\mathrm GeV}$, leading to possible signals of PS breaking expected in the not too distant future.

We do not assume any specific scale. The compactification scale and the PS breaking scale may be in any range from $10^6 \ {\mathrm GeV}- 10^{18}\ {\mathrm GeV}$, the lower they are, the stronger experimental signatures they would generate. The compactification scale is the unification scale and defined by the size of the extra dimensions. The PS breaking scale is defined by an arbitrary real constant multiplying the compactification scale (from the Wilson line), which can be arbitrarily small. 

It also turns out that this pattern of scales is also desirable from the point of view of the cosmological implications of the model. R-parity violating terms are very strongly constrained by matter-antimatter asymmetry, since they can in principle wash out any asymmetries from earlier cosmological epochs. This is in fact the case when the asymmetry arises above the SUSY breaking scale. In the present model, since it is assumed that the SUSY breaking scale is above the Pati-Salam 
breaking scale,  the stringent constraints on such parameters are automatically avoided~\cite{Mohapatra:2015fua}.

Along similar lines, the model does not lead to proton decay. This is due to the fact that baryon parity is automatic (where all quark superfields change sign while others remain the same) and the R-parity violating term in~\eqref{eq:RPV-term} does not involve baryon number. As discussed in~\cite{Mohapatra:2015fua}, this is reminiscent of models with spontaneous R symmetry breaking.

Although in the present work no attempt has been made to include supergravity, given that R-parity is broken, a possible candidate for dark matter is the gravitino, provided its lifetime is large enough. Since the gravitino decay is associated with the gravitational constant, it can be very weak, leading to a lifetime much longer than the age of the universe~\cite{Mohapatra:2015fua}.

One can see that the EPS model leads to a very distinctive scenario with possible phenomenological signatures of PS breaking (if they happen close to  $10^6$ GeV), namely $K_L\rightarrow \mu e$,
$n-\overline{n}$ oscillations and $B_{d,s}\rightarrow \mu e$~\cite{Bandyopadhyay:2015fka},
and other low energy 
consequences of the right-handed sneutrino VEVs arising from R-parity violation~\cite{FileviezPerez:2008sx,Mohapatra:2015fua,FileviezPerez:2012mj}.

\section{Conclusion}
\label{sec:conclusion}

This paper investigated the extraordinarily elegant hypothesis
that the three families of quarks and leptons may be unified, together 
with the Higgs and gauge fields of the Standard Model (SM), into a single ``particle'', namely the 
$\textbf{248}$ vector superfield of 
a ten-dimensional $E_8$ super Yang Mills (SYM) theory.
There are no free coupling constants beyond the unique gauge coupling.
Although this idea was proposed some time ago~\cite{Olive:1982ai}, it was never developed
into realistic model, and the goal of the present paper is to make progress in that direction.
Although the theory is necessarily formulated in 10d, it is based on field theory and point particles
rather than string theory, and therefore gravity is ignored in this approach.

Towards a realistic model along these lines,
a class of orbifoldings based on 
$T^6/(\mathbb{Z}_N\times \mathbb{Z}_M)$ have been proposed and explored, that can in principle break $E_8$ SYM 
down to the SM gauge group, embedded in a larger group such as $E_6$, $SO(10)$ or $SU(5)$, together with other gauge group factors which can be broken by Wilson lines.
A discussion has been presented for some examples of $E_8$ breaking for various values of $N,M$,
including: $E_6\times SU(3)_{f}$ from $T^6/\mathbb{Z}_3$; 
E$_6$SSM from $T^6/(\mathbb{Z}_3\times \mathbb{Z}'_3)$;
$SO(10)$ from $T^6/\mathbb{Z}_6$ and $SU(5)$ from a different $T^6/\mathbb{Z}_6$.
Also, a general search for the 
SM from $T^6/(\mathbb{Z}_N\times \mathbb{Z}_M)$ was performed, that led to the result that
no solutions of this kind exist, apart from those which include a
Pati-Salam (PS) gauge symmetry.

The most promising example seems to be an Exceptional Pati-Salam (EPS)
model based on $T^6/(\mathbb{Z}_6\times \mathbb{Z}_2)$ which breaks the 10d $E_8$ SYM down to 
a SUSY PS gauge group in 4d, together with other gauge group factors, which may be broken to the SM gauge group
via Wilson lines and other VEVs. In particular,
the RH sneutrinos are assumed to obtain large VEVs through a Wilson line,
generating the flavour structure in the fermion mass matrices and a mixing between the lepton doublet and the down type Higgs. This distinguishes between charged lepton and down quark masses, giving Majorana masses to the right handed neutrino and breaking the PS gauge group down to that of the SM together with $U(1)_N$, which may be broken at some other scale by other singlet VEVs.
This implies R-parity violation, and experimental signatures
associated with a low PS gauge group breaking scale.
In the EPS model, 
all the SM gauge fields, fermions, Higgs fields plus right handed neutrinos come from the same original 
$\textbf{248}$ vector superfield of $E_8$ SYM, although anomaly cancellation requires additional states.
This model raises additional phenomenological questions related to unification, the mass spectrum of states, the various thresholds in the theory, the details of the fermion mass spectrum, and so on. It would be interesting to return to these questions in a future publication.

\subsection*{Acknowledgements}

S.\,F.\,K. acknowledges the STFC Consolidated Grant ST/L000296/1 and the European Union's Horizon 2020 Research and Innovation programme under Marie Sk\l{}odowska-Curie grant agreements Elusives ITN No.\ 674896 and InvisiblesPlus RISE No.\ 690575. A.\,A. acknowledges support form CONACYT-SNI (M\'exico). We thank Patrick K. S. Vaudrevange for his insightful and crucial remarks 
and for carefully reading the final version of the manuscript.

\appendix

\section{Pati-Salam symmetry from a general scan}
\label{app:orb}

The general scan for the search for the SM gauge group included
an orbifold $T^6/(\mathbb{Z}_6\times \mathbb{Z}_2)$ defined by
\begin{equation}
\mathbb{Z}_6:\ \phi \to e^{2i\pi q_{X'}/6}\phi,\ \ \ \mathbb{Z}_2:\ \phi\to e^{2i\pi q_{Y}/2}\phi,
\end{equation}
which are applied respectively as
\begin{equation}\begin{split}
(x,z_1,z_2,z_3) &\sim (x,\alpha^2 z_1,\alpha^5 z_2, \alpha^5 z_3),\\
(x,z_1,z_2,z_3) &\sim (x, -z_1, -z_2, (-1)^2 z_3),
\end{split}\end{equation}
where $\alpha=e^{2i\pi/6}$ and $-1=e^{2i\pi/2}$.

This orbifold leaves the zero modes,
in the colour-coded notation of Eq.\ref{eq:e8sm},
which transform under $SU(3)_C\times SU(2)_L \times U(1)_Y\times U(1)_X\times U(1)_{X'}\times SU(3)_F$ as:
\begin{equation}\begin{split}
V_\mu &: \textcolor{magenta}{(\textbf{8},\textbf{1},0,0,0,\textbf{1})}+\textcolor{magenta}{(\textbf{1},\textbf{3},0,0,0,\textbf{1})}+\textcolor{magenta}{(\textbf{1},\textbf{1},0,0,0,\textbf{8})}\\
&\quad+\textcolor{magenta}{(\textbf{1},\textbf{1},0,0,0,\textbf{1})}+\textcolor{magenta}{(\textbf{1},\textbf{1},0,0,0,\textbf{1})}+\textcolor{magenta}{(\textbf{1},\textbf{1},0,0,0,\textbf{1})}\\
&\quad+(\textbf{1},\textbf{1},6,4,0,\textbf{1})+(\bar{\textbf{3}},\textbf{1},-4,4,0,\textbf{1})+(\textbf{1},\textbf{1},-6,-4,0,\textbf{1})+(\textbf{3},\textbf{1},4,-4,0,\textbf{1})\\
\phi_1&:\textcolor{ForestGreen}{(\textbf{1},\textbf{2},3,2,-2,\textbf{3})}+\textcolor{ForestGreen}{(\textbf{1},\textbf{2},-3,-2,-2,\textbf{3})},\\
\phi_2&:\textcolor{blue}{(\textbf{3},\textbf{2},1,-1,1,\textbf{3})}+\textcolor{blue}{(\textbf{1},\textbf{2},-3,3,1,\textbf{3})},\\
\phi_3&: \textcolor{blue}{(\textbf{1},\textbf{1},6,-1,1,\textbf{3})}+ \textcolor{blue}{(\bar{\textbf{3}},\textbf{1},-4,-1,1,\textbf{3})}+\textcolor{blue}{(\bar{\textbf{3}},\textbf{1},2,3,1,\textbf{3})}+\textcolor{violet}{(\textbf{1},\textbf{1},0,-5,1,\textbf{3})},\\
\end{split}\end{equation}
where the electric charge generator is given by $Q=T_{3L}+Y/6$ in our normalisation.

The chiral multiplets can be written as
\begin{equation}\begin{split}
\phi_1&: \textcolor{ForestGreen}{h_u}+\textcolor{ForestGreen}{h_d},\\
\phi_2&:\textcolor{blue}{Q}+\textcolor{blue}{L},\\
\phi_3&:\textcolor{blue}{e^c}+\textcolor{blue}{u^c}+\textcolor{blue}{d^c}+\textcolor{violet}{\nu^c}.
\end{split}\end{equation}

Note that the $V_\mu$ contains some fields seemingly not in the adjoint representation. 
This means that symmetry breaking is really not to the SM, but instead corresponds to an enhanced PS gauge group
$SU(4)_{PS}\times SU(2)_L \times  SU(2)_R$, together with the other gauge group factors.
The group embedding is $SU(4)_{PS}\rightarrow SU(3)_C\times U(1)_{B-L}$ where $\textbf{4}\rightarrow (\textbf{3},1/3)
+ (\textbf{1},-1)$ and the hypercharge generator is given by $Y/6=T_{3R}+(B-L)/2$ in our normalisation.

\section{Non-SUSY Pati-Salam Model}
\label{nonsusy}
Consider a Pati-Salam model broken SUSY in an alternative orbifold
$T^6/(\mathbb{Z}_6\times \mathbb{Z}_2\times\mathbb{Z}_2)$.
The orbifolding $\mathbb{Z}_6\times \mathbb{Z}_2\in SU(3)_R$ breaks the extended SUSY. It preserves the remaining $U(1)_R$
There is an available extra orbifolding $\mathbb{Z_L}\in U(1)_R$ that would break the usual remaining $U(1)_R$ and therefore breaking simple SUSY.

In the previous $E_8\to SU(4)_{PS}\times SU(2)_L\times SU(2)_R\times SU(3)_F\times U(1)_{X'}$ orbifold breakings one can see that $\phi_1$ decomposes into the Higgses and flavons, that must be scalars, the $\phi_2$ into right handed fermions and $\phi_3$ left handed fermions. They are already separated so one can easily impose the extra $\mathbb{Z}_2$ orbifolding
\begin{equation}\begin{split}
(x,z_1,z_2,z_3)&\sim (x, z_1^*,- z_2^*, - z_3^*),
\end{split}\end{equation}
that breaks the remaining SUSY and $U(1)_R\to \mathbb{Z}_2^R$ respecting the usual R parity. This orbifolding leaves the zero modes as
\begin{equation}\begin{split}
V_\mu &: {\rm real\ vector}:\ \textcolor{magenta}{G_\mu}+\textcolor{magenta}{W^L_\mu}+\textcolor{magenta}{W^R_\mu}+\textcolor{magenta}{Z'_\mu}+\textcolor{magenta}{F_\mu},\\
\phi_1&: {\rm complex\ scalar}:\ \textcolor{ForestGreen}{h_u}+\textcolor{ForestGreen}{h_d},\\
\phi_2&: {\rm Weyl\ fermion}: \ \textcolor{blue}{Q}+\textcolor{blue}{L},\\
\phi_3&:{\rm Weyl\ fermion}: \ \textcolor{blue}{e^c}+\textcolor{blue}{u^c}+\textcolor{blue}{d^c}+\textcolor{violet}{\nu^c},\\
\end{split}\end{equation}
which is exactly the representations needed. 

Therefore the orbifold $T^6/(\mathbb{Z}_6\times \mathbb{Z}_2\times\mathbb{Z}_2)$ can break into Pati-Salam without SUSY. This SUSY breaking happens at the compactification scale which is identified with the GUT scale. 
Although this is an interesting possibility, it was not pursued in the main text.

\end{document}